\documentclass[journal]{IEEEtran}

\usepackage{xcolor,soul,framed} 

\colorlet{shadecolor}{yellow}
\usepackage[pdftex]{graphicx}

\graphicspath{{../pdf/}{../jpeg/}}
\DeclareGraphicsExtensions{.pdf,.jpeg,.png}

\usepackage[cmex10]{amsmath}
\usepackage{array}
\usepackage{mdwmath}
\usepackage{mdwtab}
\usepackage{eqparbox}
\usepackage{url}

\usepackage{cite}
\usepackage{booktabs} 
\usepackage{multirow}
\usepackage{multicol}
\usepackage{amsfonts}
\usepackage{float}
\usepackage{enumerate} 

\usepackage{amssymb}
\usepackage{pifont}
\newcommand{\cmark}{\ding{51}}%
\newcommand{\xmark}{\ding{55}}%

\begin{document}
    \title{USEV: Universal Speaker Extraction with \\ Visual Cue}
    \author{Zexu~Pan,~\IEEEmembership{Member,~IEEE},
      Meng~Ge,
      and~Haizhou~Li,~\IEEEmembership{Fellow,~IEEE}

    \thanks{This research is supported by  
    The Science and Engineering Research Council, Agency for Science, Technology and Research (A*STAR), Singapore, through the National Robotics Program under Human-Robot Interaction Phase 1 (Grant No. 192 25 00054), 
    by the Deutsche Forschungsgemeinschaft (DFG, German Research Foundation) under Germany's Excellence Strategy (University Allowance, EXC 2077, University of Bremen), 
    and by the internal project of Shenzhen Research Institute of Big Data under the Grant No. T00120220002, the Guangdong Provincial Key Laboratory of Big Data Computing under the Grant No. B10120210117-KP02, The Chinese University of Hong Kong, Shenzhen (CUHK-Shenzhen), the University Development Fund, CUHK-Shenzhen, under the Grant No. UDF01002333 and UF02002333.
    }
    \thanks{Zexu Pan is with the Integrative Sciences and Engineering Programme, the Institute of Data Science, the Department of Electrical and Computer Engineering, National University of Singapore, 119077 Singapore (e-mail: pan\_zexu@u.nus.edu).}
    \thanks{Meng Ge is with the Department of Electrical and Computer Engineering, National University of Singapore, 119077 Singapore, and the Tianjin Key Laboratory of Cognitive Computing and Application, College of Intelligence and Computing, Tianjin University, 300072 Tianjin, China (e-mail: gemeng@tju.edu.cn).}
    \thanks{Haizhou Li is with the School of Data Science, the Chinese University of Hong Kong, Shenzhen, 518172 China, the University of Bremen, 28359 Germany, and Kriston AI, Xiamen, China (e-mail: haizhouli@cuhk.edu.cn).}
    }

\maketitle

\begin{abstract}
A speaker extraction algorithm seeks to extract the target speaker's speech from a multi-talker speech mixture. The prior studies focus mostly on speaker extraction from a highly overlapped multi-talker speech mixture. However, the target-interference speaker overlapping ratios could vary over a wide range from 0\% to 100\% in natural speech communication, furthermore, the target speaker could be absent in the speech mixture, the speech mixtures in such universal multi-talker scenarios are described as \textit{general speech mixtures}. The speaker extraction algorithm requires an auxiliary reference, such as a video recording or a pre-recorded speech, to form top-down auditory attention on the target speaker. We advocate that a visual cue, i.e., lip movement, is more informative than an audio cue, i.e., pre-recorded speech, to serve as the auxiliary reference for speaker extraction in disentangling the target speaker from a \textit{general speech mixture}. In this paper, we propose a universal speaker extraction network with a visual cue, that works for all multi-talker scenarios. In addition, we propose a scenario-aware differentiated loss function for network training, to balance the network performance over different target-interference speaker pairing scenarios. The experimental results show that our proposed method outperforms various competitive baselines for \textit{general speech mixtures} in terms of signal fidelity.
\end{abstract}

\begin{IEEEkeywords}
Multi-modal, target speaker extraction, sparsely overlapped speech, \textit{general speech mixture}, scenario-aware differentiated loss.
\end{IEEEkeywords}

\IEEEpeerreviewmaketitle

\section{Introduction}
\label{sec:introduction}

\IEEEPARstart{A}{t} a cocktail party, human has the inherent ability to selectively listen to a speaker in the presence of interference speakers and background noises, this is also known as selective auditory attention~\cite{cherry1953some,Chenglin2020spex,vzmolikova2019speakerbeam,wang2019voicefilter,pan2021reentry}. Decades of efforts have been spent to emulate such attention with engineering solutions, which is also referred to as the \textit{cocktail party} problem~\cite{bronkhorst2000cocktail}. It is non-trivial but highly demanded in real-world applications such as hearing aids~\cite{wang2017deep}, active speaker detection and verification~\cite{tao2021someone,tao2020audio}, speaker localization~\cite{qian2021multi}, and automatic speech recognition~\cite{yue2019end}.

Speech separation represents one way to address the \textit{cocktail party} problem, which seeks to separate a multi-talker speech mixture into individual streams, each for one speaker. The traditional approaches,  e.g., non-negative matrix factorization~\cite{schmidt2006single, cichocki2006new}, factorial hidden Markov models and Gaussian mixture models~\cite{stark2010source}, and computational auditory scene analysis~\cite{lyon1983computational,wang2006computational,hu2007auditory}, explore the idea of spectro-temporal masking based on harmonic and pitch analysis to filter a speaker out from a speech mixture. The prior studies have laid the foundation for recent progress. With the advent of deep learning, speech separation has seen major progress~\cite{xu2018single,kavalerov2019universal,chen2017deep,stoller2018wave,kolbaek2017multitalker,liu2019divide,9053461,9657514,han2022dpccn,subakan2022real,taherian2022location}, even with different speaker overlapping ratios~\cite{von2019,vonneumann21_GraphPIT,von2022sasdr,li2021dual,yoshioka2018multi,chen2021continuous,wang2021count,zhang2022continuous,zhang2022all}. In a neural architecture, multiple speaker streams compete and segregate either with a masking or regression mechanism. As the number of speakers in the speech mixture is typically required in advance, speech separation has a limited scope of applications.

Speaker extraction is different from speech separation. It only extracts the speech of a target speaker from a multi-talker speech mixture, thus invariant to the number of speakers. Speaker extraction emulates a human's selective auditory attention at a \textit{cocktail party}, which typically requires an auxiliary reference, that provides the information of the target speaker~\cite{ochiai2019multimodal,sato2021multimodal,8462661,Chenglin2020spex,pan2022seg,pan2022hybrid,wu2022time}. Therefore, the quality of the auxiliary reference matters. A pre-recorded speech from the target speaker has been well studied to serve as such an auxiliary reference~\cite{Chenglin2020spex,vzmolikova2019speakerbeam,wang2019voicefilter,spex_plus2020,ge2020multi,he2020speakerfilter,xiao2019single,shi2020speaker,delcroix2020improving}, the speaker extraction algorithm is expected to extract only the speech that has a similar voice signature to the pre-recorded speech.

The state-of-the-art speaker extraction algorithms perform remarkably well on highly overlapped speech, in which the speakers overlap almost 100\%, e.g., WSJ0-2mix-extr dataset~\cite{Chenglin2020spex,hershey2016deep}. However, in natural speech communication, a target speaker may speak and pause intermittently, interspersed with the speech of interference speakers, resulting in sparsely overlapped speech, i.e., the speaker overlapping ratio is only around 20\% in meetings~\cite{ccetin2006analysis}, and several datasets have been recorded to simulate such natural conversational speech with various speaker overlapping ratios, e.g., CHiME~\cite{barker2018fifth} and LibriCSS~\cite{chen2020continuous} datasets. Speaker extraction algorithms for such sparsely overlapped speech have less been studied.

In the task of speaker extraction, the target speaker can be absent or present in a speech mixture~\cite{marvin2021,zhang2020x}. Target absent (\textit{TA}) refers to the scenario where the target speaker is quiet throughout the speech mixture clip. Target present (\textit{TP}) refers to the scenario where the target speaker is present in the speech mixture clip, and the target-interference speaker overlapping ratio could range from 0\% to 100\%. Speech mixtures at a varying speaker overlapping ratio, with either absent or present target speaker, are referred to as \textit{general speech mixtures} in this paper.

\begin{figure}
  \centering
  \includegraphics[width=0.99\linewidth]{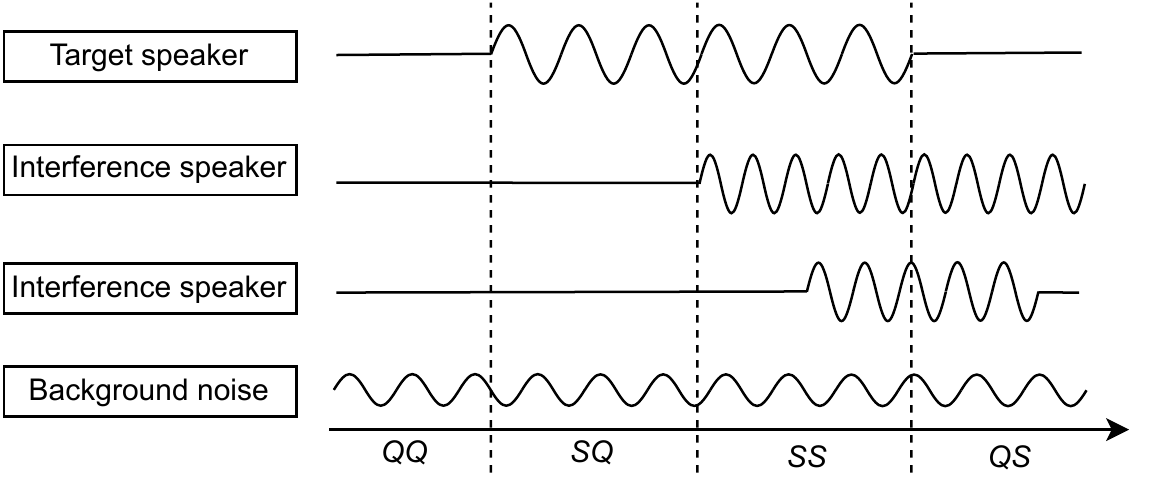}
  \caption{An illustration of the four target-interference speaker pairing scenarios, namely $QQ, SQ, SS, QS$, that can occur within a \textit{general speech mixture} clip.}
\label{fig:scenario}
\end{figure}

\begin{figure}
\begin{minipage}[b]{.99\linewidth}
  \centering
  \centerline{\includegraphics[width=\linewidth]{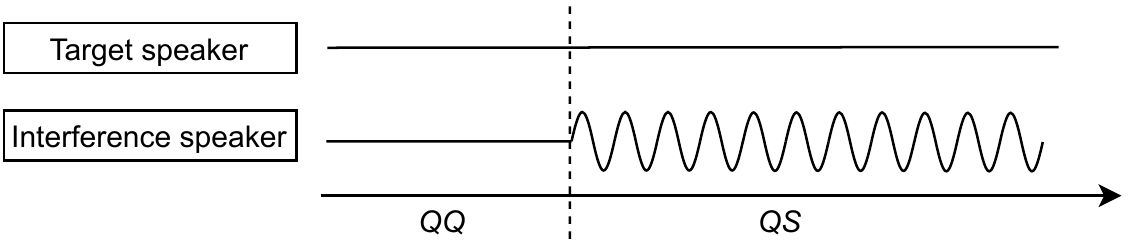}}
  \centerline{\scalebox{0.8}{(a) An \textit{TA} speech clip that consists of a mix of $QQ$ and $QS$ scenarios.}}\medskip
\end{minipage}
\hfill
\begin{minipage}[b]{0.99\linewidth}
  \centering
  \centerline{\includegraphics[width=\linewidth]{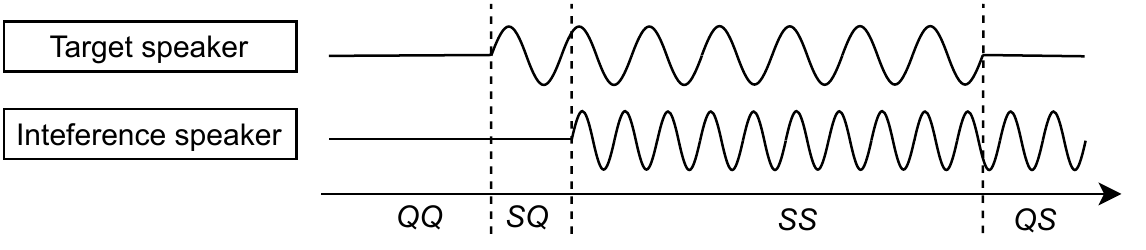}}
  \centerline{\scalebox{0.8}{(b) A \textit{TP} speech clip where target and interference speakers overlap 70\%.}}\medskip
\end{minipage}
\hfill
\begin{minipage}[b]{0.99\linewidth}
  \centering
  \centerline{\includegraphics[width=\linewidth]{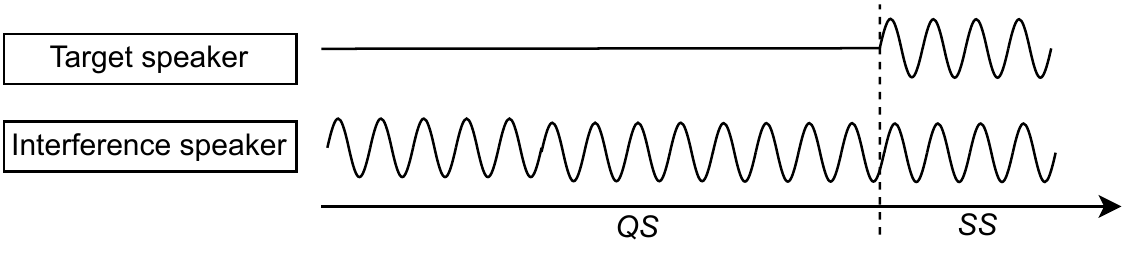}}
  \centerline{\scalebox{0.8}{(c) A \textit{TP} speech clip where target and interference speakers overlap 30\%.}}\medskip
\end{minipage}
\caption{Illustration of some speech mixture clips. (a) is an \textit{TA} example.  (b) and (c) are two \textit{TP} examples.}
\label{fig:example_utt}
\end{figure}

We argue that the \textit{TA} and \textit{TP}  categorization is too broad to be technically meaningful as far as speaker extraction is concerned, and advocate four possible target-interference speaker pairing scenarios, that could occur in a \textit{general speech mixture}, namely $QQ, SQ, SS, QS$, in Fig.~\ref{fig:scenario}. $QQ$ denotes the scenario of \textit{quiet} target speaker with \textit{quiet} interference speakers; $SQ$ denotes that of \textit{speaking} target speaker with \textit{quiet} interference speakers; $SS$ denotes that of \textit{speaking} target speaker with \textit{speaking} interference speakers; $QS$ denotes that of \textit{quiet} target speaker with \textit{speaking} interference speakers. A \textit{general speech mixture} clip in practice could be a random combination of any of the four scenarios, where $SS$ represents the scenario of highly overlapped speech in many prior studies~\cite{spex_plus2020,vzmolikova2019speakerbeam,wang2019voicefilter,hershey2016deep}, while $QQ, SQ, QS$, and the scenario transition in between are not well studied yet. In Fig.~\ref{fig:example_utt}, we illustrate some examples of the \textit{TA} and \textit{TP} speech clips. It is worth noting that the \textit{TA} speech clips can be in either $QQ$ or $QS$ scenarios, or both. But $QQ$ and $QS$ scenarios can also possibly take place in the \textit{TP} speech clips.

A speaker extraction algorithm trained on highly overlapped speech may falsely generate undesired output when the target speaker is quiet~\cite{marvin2021}, which is called the `false-extraction' problem. By simply fine-tuning a system on speech mixture at a variety of overlapping ratios, one may address the `false-extraction' problem to some extent. However, speech mixtures of different overlapping ratios have different training expectations and difficulties for each target-interference speaker pairing scenario, thus a differentiated training strategy is needed. For example, for a highly overlapped speech mixture with a talkative target speaker, the algorithm just needs to focus on speech disentanglement between speakers, otherwise, the algorithm needs to do well in verifying whether the target speaker is active as well.

Humans perceive the world through various simultaneous sensory systems~\cite{smith2005development}. We attend to a speaker not only by reference to a registered voice signature but also through other means such as observing the lip movement~\cite{afouras2018conversation} or understanding the contextual relevance~\cite{Li2020,li2022vcse}. Neuroscience studies suggest that hearing is improved by observing the lip movement in a conversation~\cite{golumbic2013visual,crosse2016eye}, leveraging on the temporal correlation between visemes and speech~\cite{edelman1987neural}. A viseme is a basic unit of visual speech, which corresponds to a set of phonemes for acoustic speech~\cite{bear2017phoneme,massaro2014speech}.

Although visemes and phonemes do not have a one-to-one correspondence, visemes do provide a fine-grained cue about underlying phonetic units being spoken~\cite{michelsanti2021overview}. Therefore, visemes have been widely adopted as the auxiliary reference for speaker extraction, which performs exceptionally well to disentangle the target speaker from high overlapped speech~\cite{ephrat2018looking,wu2019time,ochiai2019multimodal,pan2020muse}. Visemes also provide a high-level cue that discriminates between speech and non-speech signals~\cite{michelsanti2021overview}, thus possibly alleviating the `false-extraction' problem. We advocate that visemes are more informative than a pre-recorded speech to discriminate the speaking status of the target speaker at frame-level for a \textit{general speech mixture}, thus serving as a robust auxiliary reference for speaker extraction algorithms. Next, we summarize our contributions of this work.
 
\begin{enumerate}

    \item We address a unique research problem, i.e., speaker extraction from a \textit{general speech mixture} with visual auxiliary reference, to deal with all multi-talker scenarios with one universal solution. This study brings audio-visual speaker extraction a step closer to solving the \textit{cocktail party} problem in real-world applications.
    
    \item We propose to categorize a \textit{general speech mixture} into four different target-interference speaker pairing scenarios, and devise a scenario-aware differentiated loss function, to moderate the training of the four scenarios. We confirm the effectiveness through experiments on various speaker extraction models.
    
    \item We evaluate the state-of-the-art speaker extraction algorithms on \textit{general speech mixtures} and confirm the effectiveness of visual reference over speech reference. We also study the use of the dual-path recurrent neural network architecture in the speaker extractor, instead of the dilated temporal convolutional neural network, to capture global temporal dependencies in speech.
    
\end{enumerate}

The rest of the paper is organized as follows. In Section~\ref{sec:related_works}, we discuss related work. In Section~\ref{sec:methodology}, we formulate the proposed method. In Section~\ref{sec:experiment}, we present the experimental setup. In Section~\ref{sec:results}, we report the experimental results. Finally, we conclude the study in Section~\ref{sec:conclusion}.

\section{Related work}
\label{sec:related_works}
The speaker extraction algorithm usually requires an auxiliary reference to form the top-down attention on the target speaker. Either an auditory or a visual cue may serve as a such auxiliary reference. Next, we discuss the prior studies for three related tasks to set the stage for our study.

\subsection{Highly overlapped speech mixture with audio cue}
Each speaker has a unique voice signature, which can be characterized by a speaker embedding, such as i-vector~\cite{dehak2010front}, x-vector~\cite{snyder2016deep}, and d-vector~\cite{wan2018generalized}. Given a speech sample, the speaker embedding can be derived by a speaker encoder that is trained with a speaker recognition task~\cite{liu2020speaker}. 

VoiceFilter~\cite{wang2019voicefilter} employs the d-vector encoded from a pre-recorded speech, that serves as an audio cue, to form the top-down attention for speaker extraction. The study explores the discriminative power of the d-vector to find the speech track in the multi-talker speech that has a similar voice signature to the d-vector, and extracts that speech track. The idea of speaker embedding is extended to the context of speaker extraction, where the speaker encoder is jointly trained with the speaker extractor network in a multi-task learning framework~\cite{vzmolikova2017learning,Chenglin2020spex,spex_plus2020}. The speaker extractor network is trained by a signal reconstruction loss while the speaker encoder is trained with a speaker recognition loss. In this way, the resulting speaker embedding is not only discriminative between speakers, but also optimized for speaker extraction. 

These prior studies assumed that speakers highly overlap in the multi-talker speech, and the target speaker is always present. The case of highly overlapped speech puts the speaker extraction algorithm in a stress test. However, it only represents one of many possible multi-talker scenarios. There are other scenarios, for example, the speaker of interest is absent, or the speaker of interest is the only speaker in the speech. To work in real-world applications, a speaker extraction model needs to perform across all scenarios.

\subsection{Absent target speaker speech mixture with audio cue}   
In natural speech communication, interlocutors typically take turns to speak, resulting in speech mixtures of varying extents of overlapping. In particular, a speaker of interest could remain quiet throughout while some others are speaking, resulting in \textit{TA} mixture clips. There have been studies to include \textit{TA} clips into the training data to overcome `false-extraction'~\cite{zhang2020x,marvin2021}. It is noted that the commonly used objective function scale-invariant signal-to-noise ratio (SI-SDR)~\cite{le2019sdr} or signal-to-noise ratio (SDR) is undefined, or defined as a constant, for \textit{TA} clips, It was proposed~\cite{zhang2020x,fuss2021,marvin2021,vonneumann21_GraphPIT} to minimize the energy of \textit{TA} clips instead, at the same time maximizing the SI-SDR or SDR for \textit{TP} clips. We refer to such a loss as the scenario-aware uniform loss in this paper, it is scenario-aware because it differentiates \textit{TA} with \textit{TP} clips, it is uniform as it applies a single loss function to each of an entire clip. These prior studies~\cite{zhang2020x,marvin2021} focus on the case in which the target speaker is either absent or highly overlapped with the interference speaker in a speech clip. We consider the more realistic \textit{general speech mixtures} in this paper as illustrated in Fig.~\ref{fig:scenario}, where sparsely overlapped speech is included. 

A recent speech separation study~\cite{von2022sasdr} proposed the source-aggregated SDR, which computes the mean over the SDR values of each output channel, this works as long as the target speaker is not absent in at least one output channel. Other studies~\cite{zhang2022all,zhang2022continuous} used the mean squared error loss on the estimated mask or spectrogram to avoid the absent target speaker problem, while it remains unclear whether this is better as it is not directly optimizing the signal quality of the extracted speech waveform. In another study of the VAD-SE network~\cite{lin2021sparsely}, a target speaker voice activity detection (VAD) module and a speaker extraction network~\cite{lin2021sparsely} are jointly trained. During run-time inference, the VAD mutes the speaker extraction output when the target speaker is quiet. In other words, the speaker extraction performance highly depends on the VAD accuracy.

Both the scenario-aware uniform loss~\cite{zhang2020x,marvin2021,fuss2021,vonneumann21_GraphPIT} and the VAD-SE~\cite{lin2021sparsely} network use an audio cue, i.e., a pre-recorded speech, as the auxiliary reference. However, for a \textit{general speech mixture}, the audio cue is not only used for disentanglement but also used to verify whether the target speaker is speaking or quiet for every frame, which is difficult. In this paper, instead of an audio cue, we explore the use of a visual cue, i.e., lip movement, that doesn't require the pre-recording of speech for the speaker extraction task. We believe the visual cue is a more direct cue verifying the speaking status of the target speaker compared to the audio cue.

\subsection{Highly overlapped speech mixture with visual cue} 
In the \textit{cocktail party} problem, the visual reference, if present, is not corrupted by either acoustic noise or interference speech, which can be employed to form the top-down attention on the target speaker. The visual reference usually takes the form of a single face image from the target speaker, or a sequence of lip images of the target speaker that is synchronized with the speech.

FaceFilter~\cite{chung2020facefilter} makes use of a single face image and explores the general relationship between the facial appearance and the voice signature, such as age, gender, and ethnicity. The network encodes the face image into a speaker embedding to form the attention on the target speaker.

Many audio-visual speaker extraction algorithms use the visemes encoded from a sequence of lip images of the target speaker as the visual cue. The conversation~\cite{afouras2018conversation}, and Time-domain speaker extraction network (TDSE)~\cite{wu2019time} are such examples, where they pre-train a visual encoder on a visual speech recognition task~\cite{stafylakis2017combining}. The visual encoder encodes a sequence of lip images, that are synchronized with the target's speech,  into a sequence of lip embeddings, also known as visemes. The lip embeddings, that are time-aligned with speech frames, are then used to model the temporal synchronization and interaction between the lip movement and speech.

The visemes are shown effective in disentangling a target speaker from highly overlapped speech, but have not been studied for the \textit{general speech mixtures}. In this paper, we employ the visemes to form the top-down attention on the target speaker.

\begin{figure*}[th]
  \centering
  \includegraphics[width=\linewidth]{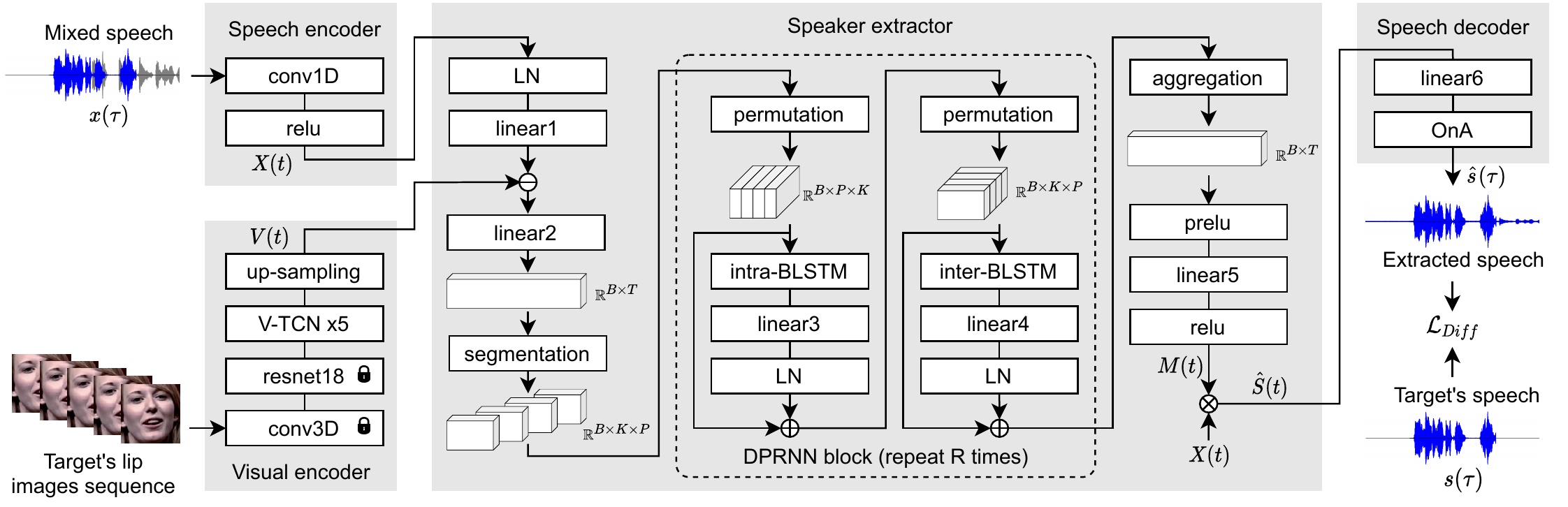}
  \caption{The proposed universal speaker extraction network, or USEV network in short. It consists of a speech encoder, a visual encoder, a speaker extractor, and a speech decoder. The symbol $\ominus$ refers to frame-wise concatenation of features; the symbols $\oplus$ and $\otimes$ refer to element-wise addition and multiplication of features. The network layers are represented by rectangles, the intermediate features are represented by 3D blocks. The lock sign in the visual encoder means that the weights of the network layers are frozen during the USEV network training.}
\label{fig:network}
\end{figure*}

\section{Universal speaker extraction network with visual cue}
\label{sec:methodology}

Let $x(\tau)$ be a multi-talker speech clip, consisting of the target speaker's speech $s(\tau)$ and interference speaker's speech $b_{i}(\tau)$,
\begin{equation}
    \label{eqa:speaker_extraction}
    x(\tau) = s(\tau) + \sum_{i=1}^{I}b_{i}(\tau),
\end{equation}
where $i \in \{1,...,I\}$ denotes the index of interference speakers.
The task of target speaker extraction aims to recover $\hat{s}(\tau)$ that is close to $s(\tau)$ from $x(\tau)$. 

\subsection{Network architecture}
We now formulate a Universal Speaker Extraction network with a Visual cue, named USEV (pronounced as `use v') network, as depicted in Fig.~\ref{fig:network}. The network is universal because it is expected to perform for a speech clip with all possible target-interference speaker pairing scenario combinations.

We adopt the time-domain approach for the network design, which is originally proposed in the time-domain speech separation network (Conv-TasNet)~\cite{luo2019conv}, to avoid the phase estimation problem arising from the frequency-domain approach. The USEV network has four components: 1) The speech encoder transforms the time-domain speech samples $x(\tau)$ into a sequence of speech embeddings $X(t)$. 2) The visual encoder encodes the target's lip images sequence into a sequence of visual embeddings $V(t)$. 3) The speaker extractor estimates a mask $M(t)$, which only lets the target speaker pass in $X(t)$. 4) The speech decoder renders the masked speech embeddings $\hat{S}(t)$ into time-domain speech samples $\hat{s}(\tau)$.

\subsubsection{Speech encoder}
The speech encoder consists of a 1-dimensional (1D) convolution $conv1D$ followed by a rectified linear activation $relu$. The speech encoder behaves like a frequency analyzer to convert the time-domain speech samples $x(\tau)$ into a spectrum-like frame-based embedding sequence $X(t)$ in the latent space,
\begin{equation}
    X(t) = relu(conv1D(x(\tau),1, N, L))  \;  \in \mathbb{R}^{N\times T}
\end{equation}
where the $conv1D$ has input channel size $1$, output channel size $N$, kernel size $L$, and stride $L/2$. The output $X(t)$ is a $T$ frame embedding sequence of dimension of $N$, where $t \in \{1,...,T\}$.

\subsubsection{Visual encoder}
The visual encoder seeks to encode the target's lip images sequence into a sequence of visual embeddings $V(t) \in \mathbb{R}^{N\times T}$, representing the target speaker's visemes, and in sync with the target's speech. We design the visual encoder with a structure similar to the visual encoder in MuSE~\cite{pan2020muse}, which consists of a 3-dimensional (3D) convolution $conv3D$, an $18$ layer residual convolutional neural network $resnet18$, $5$ repeated visual temporal convolutional network $V\mbox{-}TCN$, and an up-sampling layer.

The $conv3D$ and $resnet18$ are pre-trained from visual speech recognition task~\cite{stafylakis2017combining}, that are denoted by a lock sign in Fig.~\ref{fig:network}. With the pre-trained weights fixed during speaker extraction training, we seek to retain the pre-trained knowledge to encode visemes that synchronize with the phonetic sequence of the target's speech.

\begin{figure}
  \centering
  \includegraphics[width=2.2cm]{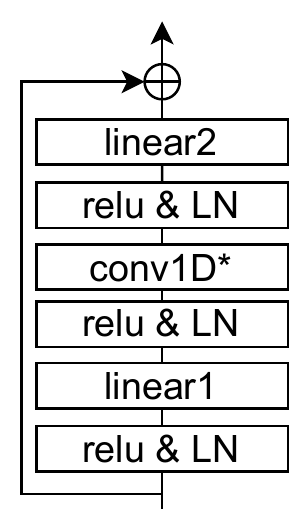}
  \caption{The architecture of the $V\mbox{-}TCN$ in the visual encoder. It consists of a linear layer $linear1$ with input and output sizes $N$ and $N*2$ respectively; a group convolution layer $conv1D^*$ with input channel size $N*2$, output channel size $N*2$, and group size $N*2$, kernel size $3$; and a linear layer $linear2$ with input and output sizes $N*2$ and $N$ respectively. Each of the $linear1$, $linear2$, and $conv1D^*$ layers are preceded with a $relu$ and a layer normalization $LN$.}
\label{fig:vtcn}
\end{figure}

The visual features extracted from the pre-trained $conv3D$ and $resnet18$ are different from the speech embeddings~\cite{pan2020multi}, that are not optimized for speaker extraction directly. We design an adaptation network with $5$ repeated $V\mbox{-}TCN$ similar to the reentry model~\cite{pan2021reentry}, with non-shared network weights, to adapt the visual embeddings towards the speaker extraction task. The architecture of a $V\mbox{-}TCN$ is shown in Fig.~\ref{fig:vtcn}. The $5$ $V\mbox{-}TCN$ is followed by an up-sampling layer to match the temporal resolution of the visual embeddings to the same as the speech embeddings $X(t)$.

\subsubsection{Speaker extractor}
In computational auditory scene analysis~\cite{brown1994computational}, the selective filter is well studied to emulate the human's selective auditory attention in attentive listening, which is usually modeled as a receptive mask. The speaker extractor in the USEV network adopts the masking method~\cite{Chenglin2020spex} to estimate a mask $M(t)$ which only lets the target speaker pass in the speech embeddings $X(t)$. The masked speech embeddings $\hat{S}(t)$ are obtained by element-wise multiplication between $X(t)$ and $M(t)$,
\begin{equation}
    \hat{S}(t) = X(t) \otimes M(t)  \;  \in \mathbb{R}^{N\times T}
\end{equation}
where $X(t)$ is the output of the speech encoder.

The speaker extractor requires a reference to form the top-down auditory attention on the target speaker. The visual embeddings $V(t)$ are trained just for that by providing the target speaker's visemes, that are synchronized with the target's speech. The inputs to the speaker extractor are the speech embeddings $X(t)$ and the visual embeddings $V(t)$. The studies on the $reentry$ model~\cite{pan2021reentry}, TDSE~\cite{wu2019time}, and MuSE~\cite{pan2020muse} suggest that, by concatenating the time-aligned visual embeddings with their corresponding speech embeddings, the speaker extractor is able to effectively estimate the mask $M(t)$. We adopt the concatenation approach at the start of the speaker extractor.

Before the concatenation operation, $X(t)$ is passed through a layer normalization $LN$, followed by a bottleneck linear layer $linear1$~\cite{luo2019conv} with input and output sizes $N$ and $B$ respectively. The concatenated embeddings are passed through a linear projection layer $linear2$ with $B+N$ and $B$  as the input and output sizes respectively.

In view that a speech utterance is usually encoded into a long sequence of speech embeddings. It is important to effectively model the long-term dependencies in the mask estimation process. Dilated temporal convolutional neural network (TCN) with a large receptive field has been widely used, such as WaveNet~\cite{oord2016wavenet}, Conv-TasNet~\cite{luo2019conv}, and TDSE~\cite{wu2019time}. However, the TCN has a fixed receptive field, and thus has difficulty in learning the long-term dependencies. 

DPRNN~\cite{luo2020dual} is a dual-path recurrent neural network for speech separation, which has a dynamic receptive field to capture global dependencies. DPRNN segments a long sequence of embeddings into short chunks and applies intra- and inter-chunk operations with an interlacing structure. In view of its success, we use the dual-path structure~\cite{luo2020dual} in the context of audio-visual speaker extraction. which consists of a segmentation layer, $R$ repeated DPRNN blocks, and an aggregation layer in the speaker extractor. 

After the linear projection layer $linear2$, the 2-dimensional (2D) vector is passed through a segmentation layer, which is segmented into chunks with a window size $K$ and hop size $K/2$. $K$ is selected such that $K \approx \sqrt{2T}$~\cite{luo2020dual}. The resultant $P$ chunks are concatenated to form a 3D tensor, where $P=\frac{2\times T}{K}+1$.

After segmentation, the 3D tensor is passed through $R$ repeated DPRNN blocks with non-shared network weights, each with a chain of intra-chuck processing and inter-chuck processing with residual connections. The intra-chunk processing has an intra-bidirectional long short-term memory \textit{intra-BLSTM}, with input and hidden size $B$ and $B\times 2$ respectively, which is applied to the intra-chunk sequence ($K$ dimension) of the 3D tensor. The \textit{intra-BLSTM} is followed by a linear layer $linear3$ with input and output sizes $B\times 4$ and $B$ respectively, and a $LN$. Similar to the intra-chunk processing, the inter-chuck has a similar network architecture and parameters, except that the \textit{inter-BLSTM} is applied to the inter-chunk sequence ($P$ dimension) of the 3D tensor.

The aggregation layer is an inverse operation of the segmentation layer, which transforms the 3D tensor back to 2D. It is followed by a parametric rectified linear activation $prelu$, a linear layer $linear5$ with input and output sizes $B$ and $N$, and a $relu$ layer. 

\subsubsection{Speech decoder}
The speech decoder renders the masked speech embeddings $\hat{S}(t)$ into time-domain speech samples $\hat{s}(t)$. It consists of a linear layer $linear6$ and an overlap-and-add operation $OnA$,
\begin{equation}
    \hat{s}(\tau) = OnA(Linear6(\hat{S}(t), N, L), L/2)
\end{equation}
where the $OnA$ operation has a frame shift of $L/2$.

\begin{figure}
  \centering
  \includegraphics[width=0.99\linewidth]{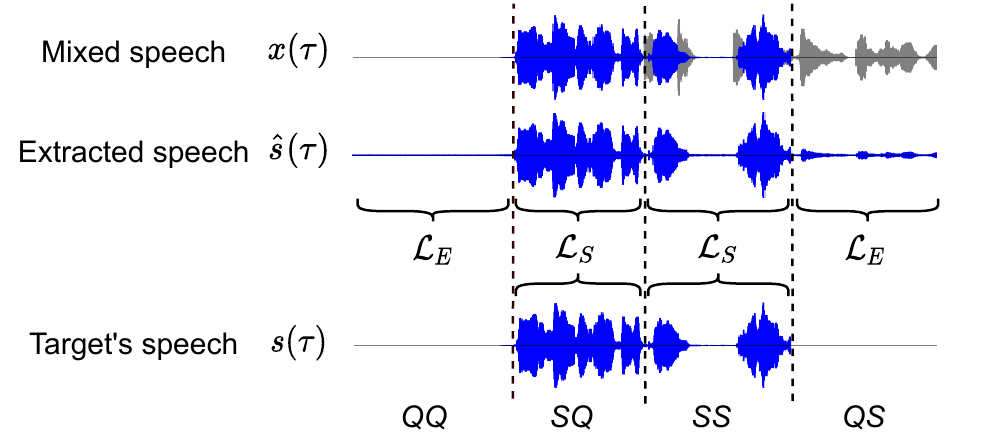}
  \caption{An illustration of the scenario-aware differentiated loss. We segment an extracted speech clip according to the different target-interference speaker pairing scenarios ($QQ, SQ, SS, QS$) and apply separate loss functions for different segments.}
\label{fig:loss}
\end{figure}

\subsection{Scenario-aware differentiated loss}
In time-domain source separation or extraction algorithms, SDR is widely used as the objective function~\cite{zeghidour2020wavesplit,delcroix2022soundbeam,wisdom2020unsupervised}, which performs very well in the case of highly overlapped sound sources. However, for \textit{TA} mixture clips, the SDR is a constant value. The scenario-aware uniform loss studies~\cite{zhang2020x,fuss2021,marvin2021,vonneumann21_GraphPIT}, which the loss function is shown in Eq.~(\ref{eqa:uniform}), point to a direction that, by adding an $\epsilon$ at the numerator, the SDR loss reduces to a form that minimizes the energy of the \textit{TA} speech clips, at the same time, still maximizes the SDR for \textit{TP} speech clips.

\begin{equation}
    \mathcal{L}_{Uni} = - 10 \log_{10} ( \frac{||s||^2 + \epsilon }{||\hat{s} - s||^2  + \epsilon} )
    \label{eqa:uniform}
\end{equation}
where $\epsilon$ is a small value of $1e^{-8}$. We omit the subscript $(\tau)$ for the clean $s(\tau)$ and extracted speech $\hat{s}(\tau)$ in the formula for brevity.

In reality, a \textit{TP} speech clip could be a combination of $QQ, SQ, SS, QS$ scenario segments. Such a multi-scenario clip is called a heterogeneous clip, as shown in Fig.~\ref{fig:scenario} and Fig.~\ref{fig:example_utt}, while a single scenario clip is called a homogeneous clip. It is understood that we have different expectations for each of the scenarios, e.g., low energy silence outputs for $QS$ and $QQ$ scenarios, and high SDR outputs for $SS$ and $SQ$ ones. On the other hand, the difficulty of training for different scenarios is expected to be different, e.g., $SS$ scenario is more complex than $SQ$, and $QS$ is more complex than $QQ$. Therefore, it is best that we measure the output speech quality of each scenario differently. However, the loss functions, either SDR or energy loss, are segmental in nature. A single uniform loss reflects the average loss quantity indiscriminately over the entire heterogeneous clip, that doesn't seek to optimize for each scenario, which could result in unbalanced training between scenarios. We argue that a scenario-specific loss for each scenario-homogeneous segment should be more appropriate.

Furthering the study of scenario-aware uniform loss, we propose a training strategy with a scenario-aware differentiated loss to deal with the four scenarios as summarized in Fig.~\ref{fig:loss}. During training, the speaker extraction network takes the entire speech clip as input and the scenario labels as the supervision signals. We segregate the extracted speech clip into the respective scenario segments according to the labels, i.e., $QQ, SQ, SS, QS$.

For the $SQ$ and $SS$ segments, where the target speaker is acoustically active, we apply the SDR as the loss $\mathcal{L}_{S}$ between the ground-truth and the extracted speech,

\begin{equation}
    \mathcal{L}_{S} = - 10 \log_{10} ( \frac{||s||^2}{||\hat{s} - s||^2  + \epsilon} + \epsilon)
    \label{eqa:sdr}
\end{equation}

For the $QQ$ and $QS$ segments where the target speaker is acoustically inactive, we apply the energy of the segments as the loss $\mathcal{L}_{E}$,

\begin{equation}
    \mathcal{L}_{E} = 10 \log_{10} ( ||\hat{s}||^2 + \epsilon)
    \label{eqa:energy}
\end{equation}

We define the total scenario-aware differentiated loss $\mathcal{L}_{Diff}$ as a weighted sum of the $4$ differentiated loss values as follows, 

\begin{equation}
    \mathcal{L}_{Diff} = \alpha \mathcal{L}_{E}^{QQ} + \beta \mathcal{L}_{S}^{SQ} + \gamma \mathcal{L}_{S}^{SS} + \delta \mathcal{L}_{E}^{QS} 
    \label{eqa:diff}
\end{equation}
where $\alpha, \beta, \gamma, \delta$ are the weights for the loss, also called the hyper-parameters. 

With the scenario-aware differentiated loss $\mathcal{L}_{Diff}$, we are able to optimize the network with a scenario-specific loss for each scenario-homogeneous segment, yet having an overall objective across the entire heterogeneous clip, i.e., a random combination of different scenarios. The weights in $\mathcal{L}_{Diff}$ moderate the contributions of the individual scenarios towards the overall objective.
 
One may argue that we may optimize the homogeneous clips i.e., clips of a single scenario, scenario by scenario as in~\cite{marvin2021}. It is noted that such a training strategy will not take care of the temporal transition between scenarios. The scenario-aware differentiated loss $\mathcal{L}_{Diff}$ for an entire heterogeneous clip is necessary.

The labels of $QQ, SQ, SS, QS$ scenarios and the segmentation process are only required during training, and not during inference. The training on heterogeneous clips with the proposed loss $\mathcal{L}_{Diff}$ allows for optimization of the network for unknown speaker pairing scenarios at run-time.

It is worth mentioning that we do not normalize the extracted speech during inference, because this may potentially amplify the unwanted noise in the $QQ$ and $QS$ scenarios. We choose the SDR loss over the SI-SDR loss for Eq.~\ref{eqa:sdr} in the proposed scenario-aware differentiated loss, because the SDR loss is scale-sensitive, thus keeping the extracted speech at the same scale as the ground truth clean speech. It is noted that the SI-SDR loss is scale-invariant~\cite{le2019sdr}, thus the scale of the extracted speech may run away without a proper normalization.

\subsection{Training procedure}
The overall training of the USEV network consists of three stages.
\begin{enumerate}
    \item The $conv3D$ and $resnet18$ layers in the visual encoder are pre-trained together with a LSTM back-end on a visual speech recognition tasks\footnote{The $conv3D$ and $resnet18$ structures follow~\cite{Afouras18b}, and their pre-trained weights on visual speech recognition task are taken from \url{https://github.com/lordmartian/deep_avsr}}. The $conv3D$ and $resnet18$ layers are then kept frozen during subsequent USEV network training.
    \item The entire USEV network is pre-trained on highly overlapped speech clips, with the SDR loss as shown in Eq.~(\ref{eqa:sdr}). This allows the USEV network to focus on the $SS$ scenario as it is the most difficult scenario among the four~\cite{marvin2021}.
    \item The entire USEV network is trained on the general speech mixture clips, with the scenario-aware differentiated loss as shown in Eq.~(\ref{eqa:diff}).
\end{enumerate}

\begin{table*}[th]
    \centering
    \caption{A summary of the number of speech clips for different target-interference speaker overlapping ratios, and total duration for 4 target-interference pairing scenarios in the simulated IEMOCAP-mix dataset.}
    \begin{tabular}{c||c |c |c |c |c |c |c ||c|c |c |c} 
       \toprule
    \multirow{2}*{}
    &\multirow{2}*{\textit{TA} clips}
    &\multicolumn{6}{c||}{\textit{TP} clips with different target-interference speaker overlapping ratio}
    &\multicolumn{4}{c}{Duration (hour)}  \\ 
             &   &0 \% & (0,20] \%   &(20,40] \%   &(40,60] \% &(60,80] \%   &(80,100] \% &$QQ$    &$SQ$   &$SS$   &$QS$ \\ 
        \midrule
        \# training        &20,437  &34,859  &24,256  &46,846  &73,533  &84,946  &115,123 &48.73  &83.19  &266.34 &133.41\\
        \# validation  &295   &585   &496   &1,112  &1,869   &2,129   &3,514 &0.85   &1.79   &7.53   &3.01   \\
        \# test        &169   &379   &282   &649   &1,081   &1,492   &1,948  &0.53   &1.05   &4.19   &1.74  \\
       \bottomrule
    \end{tabular}
    \label{table:iemocap}
\end{table*}

\section{Experimental Setup}
\label{sec:experiment}

\subsection{Dataset}

\subsubsection{Highly overlapped speech mixtures (VoxCeleb2-mix dataset)}
We use the VoxCeleb2~\cite{Chung18b} dataset to simulate a highly overlapped speech dataset, denoted as VoxCeleb2-mix. The dataset is used for model pre-training. 

The VoxCeleb2 dataset is an audio-visual dataset derived from YouTube videos. It has over 1 million videos from $6, 112$ celebrities. The videos are pre-processed with face detection and tracking algorithms. The resulting face tracking sequences are used as auxiliary references both for training and testing. We sample the audios at $16$ kHz, the videos are synchronized with the audios and sampled at $25$ FPS.

To create the VoxCeleb2-mix dataset, we randomly select $320,000$ videos from $3,200$ speakers in the original train set to create a training set ($160,000$ speech mixture clips), and $36,237$ videos from $118$ speakers in the original test set to form a test set ($3,000$ speech mixture clips). In either case, we only include videos that have at least $4$ seconds of duration. To simulate a highly overlapped speech clip, an interference speech is mixed with the target speech at a Signal-to-Noise ratio (SNR) randomly set between $10$ dB to $-10$ dB. Between the two mixing speech clips, the longer clip is truncated to the length of the shorter one to maximize the overlap. In addition, a noise from the WHAM! noise dataset~\cite{Wichern2019WHAM} is also mixed with the target speech at an SNR randomly set between $15$ dB to $-5$ dB.

\subsubsection{General speech mixtures (IEMOCAP-mix dataset)}
We use the Interactive Emotional Dyadic Motion Capture (IEMOCAP)~\cite{busso2008iemocap} dataset to simulate the \textit{general speech mixture} clips, named IEMOCAP-mix dataset. It is used to train and evaluate the contrastive baselines and the proposed USEV network. 

The IEMOCAP dataset is an acted, multi-modal dataset. It consists of $12$ hours of $150$ dyadic conversations, each conversation has $2$ speakers, with a total of $10$ speakers in the dataset. The speakers' faces are always visible throughout, either speaking or not. We sample the audios at $16$ kHz, the videos are synchronized with the audios and sampled at $25$ FPS. 

We first obtain video clips with clean utterances from the conversations. The utterances are then used to simulate the \textit{general speech mixture} clips that cover all scenarios as described in Fig.~\ref{fig:scenario}. 

From the original IEMOCAP dataset, we detect and track the faces in the videos following the VoxCeleb2 procedure~\cite{Chung18b}, and obtain $300$ face tracking videos. Based on the available speaker diarization label in the IEMOCAP dataset, we mute the audio when the face in the video is not speaking. In this way, each face tracking video consists of alternating quiet and speech segments, where a speech segment is always associated with the face in the video. We randomly select $240$, $30$, and $30$ face tracking videos to simulate IEMOCAP-mix training, validation, and test sets respectively. A speaker may appear in multiple sets, but the speech does not. We split a face tracking video into multiple utterances by using a random quiet position between adjacent utterances as the delimiter.

To simulate the \textit{general speech mixture} clips, we drop the utterances that are shorter than $3$ seconds. We use a random segment from a target speaker's utterance, and mix it with random segments from $1$ or $2$ interference speaker's utterances at an SNR ratio between $10$ dB to $-10$ dB. Each speech clip could contain speech under one or more of the four scenarios as described in Fig.~\ref{fig:scenario}. In this way, the dataset covers all four scenarios. We simulate $400,000$, $10,000$, and $6,000$ speech mixture clips for IEMOCAP-mix training, validation and test sets. The average length of the speech mixture clips is $5$ seconds.

Besides the clean IEMOCAP-mix dataset simulated, we simulated a noisy version of the IEMOCAP-mix dataset, by adding noise from the WHAM! noise dataset~\cite{Wichern2019WHAM} to the clean version of the IEMOCAP-mix dataset with an SNR randomly set between $15$ dB to $-5$ dB.

\subsubsection{Composition of IEMOCAP-mix}

A \textit{general speech mixture} contains speech under one or more of the four scenarios, namely $QQ, SQ, SS, QS$. We record the start and endpoints of the scenarios in each speech clip. With such scenario labels, the training data are ready for network training with the scenario-aware differentiated loss, while the test data can be used for reporting evaluation performance by scenarios. We report the total duration in hours for the $4$ scenarios in the last four columns of Table~\ref{table:iemocap}.

As a \textit{general speech mixture} could also be categorized as \textit{TA} and \textit{TP} speech clips, we report the total number of \textit{TP} speech clips in groups by the target-interference speaker overlapping ratio, i.e., 0\%, (0,20]\%, (20,40]\%, (40,60]\%, (60,80]\%, and (80,100]\% in Table~\ref{table:iemocap}. The target-interference speaker overlapping ratio is defined as the ratio of the duration of the overlapped segment $SS$ to the total duration of non-silence speech $SS$+$SQ$+$QS$, i.e., ($SS$ duration)/($SS$+$SQ$+$QS$ duration). We also report the total number of \textit{TA} speech clips, the \textit{TA} speech clips have no target speaker, therefore the target-interference speaker overlapping ratio does not apply. To simulate real-world data, we allow multiple target-interference speaker pairing scenarios to take place in a single speech clip.

\subsection{USEV and contrastive baselines}
We select two time-domain target speaker extraction networks as the contrastive baselines, namely SpEx+~\cite{spex_plus2020} and TDSE~\cite{wu2019time}. The SpEx+ network employs a pre-recorded speech as the auxiliary reference, that shows the state-of-the-art performance. The TDSE network represents the recent advances of using the target's visemes as the auxiliary reference. The USEV network is a departure from the SpEx+ and TDSE networks. It is worth noting that the original SpEx+ and TDSE networks are trained and evaluated for highly overlapped speech in the previous works, which is neither a universal nor realistic acoustic scenario. We train the SpEx+, TDSE, and USEV network on general speech mixture clips, and seek to address all possible multi-talker scenarios in \textit{cocktail party}, which represents an important step toward real-world applications.

To compare the proposed scenario-aware differentiated loss in Eq.~(\ref{eqa:diff}) with the scenario-aware uniform loss~\cite{fuss2021}, we develop contrastive baselines with the scenario-aware uniform loss function for the SpEx+, TDSE, and USEV networks respectively. 

\begin{enumerate}[(i)]
    \item \textit{SpEx+}: The SpEx+ network has a similar architecture to the USEV network, except that the SpEx+ has a speaker encoder, which encodes the audio cue, in place of a visual encoder. In addition, the speaker extractor of the SpEx+ consists of repeated stacks of TCN blocks instead of repeated DPRNN blocks in USEV. We re-implement the SpEx+ by using scenario-aware uniform loss as the training objective, which is referred to as SpEx+(S).  We also re-implement another variant of the SpEx+ network by using the scenario-aware differentiated loss as the training objective, which is referred to as SpEx+(D). Both the SpEx+(S) and SpEx+(D) serve as the contrastive baselines of the USEV network.
    \item \textit{TDSE}: The TDSE network has a similar architecture to the USEV network, except that the speaker extractor in the TDSE consists of repeated stacks of TCN blocks instead of repeated DPRNN blocks in USEV. We re-implement the TDSE by using scenario-aware uniform loss as the training objective, which is referred to as TDSE(S).  We also re-implement another variant of the TDSE network by using the scenario-aware differentiated loss as the training objective, which is referred to as TDSE(D).
    \item \textit{USEV}: The USEV network with the scenario-aware differentiated loss is denoted as USEV(D). We also implement a variant of the USEV network with scenario-aware uniform loss, i.e., USEV(S) for comparison. 
\end{enumerate}

\subsection{Implementation details}
The SpEx+, TDSE, and USEV networks have the same training procedure. They are pre-trained on highly overlapped speech clips first (VoxCeleb2-mix dataset), then trained and evaluated on the \textit{general speech mixture} clips (IEMOCAP-mix dataset).

For pre-training on the VoxCeleb2-mix dataset, we use the adam optimizer with an initial learning rate of $0.001$. The learning rate is decreased by $2\%$ for every epoch, we train the networks for $30$ epochs. During pre-training, the speech clips are truncated to 6 seconds to fit into the GPU memory.

For training on the IEMOCAP-mix dataset, we use the adam optimizer with an initial learning rate of $0.0001$. The learning rate is decreased by $2\%$ for every epoch, the training stops when the best validation loss does not improve for $8$ consecutive epochs. During training, the speech clips are truncated to $6$ seconds to fit into the GPU memory, during inference, the full speech clips are evaluated\footnote{The codes for the data generation and the USEV network are available at https://github.com/zexupan/USEV.}. For the SpEx+ and TDSE networks, the $L$, $B$, $N$, and $R$ are set to $40$, $256$, $256$, and $4$ according to~\cite{wu2019time}. For the USEV network, the $L$, $B$, $N$, $R$, and $K$ are set to $40$, $64$, $256$, $6$, and $100$ according to~\cite{luo2020dual}.

\subsection{Evaluation metrics}
We use SI-SDR (dB) or power (dB/s) to evaluate our proposed method, which the two metrics are shown as follows:

\begin{equation}
    SI\mbox{-}SDR =  10 \log_{10} ( \frac{||\frac{<\hat{s},s>s}{||s||^2+ \epsilon}||^2}{||\hat{s} - \frac{<\hat{s},s>s}{||s||^2  + \epsilon}||^2  + \epsilon} + \epsilon)
    \label{eqa:si-sdr}
\end{equation}

\begin{equation}
    Power = 10 \log_{10} (\frac{||\hat{s}||^2}{T_s} + \epsilon)
    \label{eqa:power}
\end{equation}
where $T_s$ is the duration of $\hat{s}$ in seconds.

\section{Results}
\label{sec:results}

We report three groups of experiments on the IEMOCAP-mix dataset. First, we empirically study the hyper-parameters settings and the training strategy, which are reported in Section~\ref{sec:results_param}. Second, we present 3 comparative studies, that are reported in Section~\ref{sec:results_compare}. We compare the visual cue with the audio cue in experiment 1,  the scenario-aware differentiated loss with the scenario-aware uniform loss in experiment 2, and the DPRNN structure with the TCN structure in experiment 3. Third, we study how the quality of the visual cue affects the speaker extraction performance of the USEV and TDSE networks, which are reported in Section~\ref{sec:results_visual}. 

\subsection{System tuning}
\label{sec:results_param}

\subsubsection{Weights for the scenario-aware differentiated loss}
We evaluate the USEV(D) on the clean and noisy IEMOCAP-mix validation set to empirically find the appropriate weights $\alpha, \beta, \gamma, \delta$ for the differentiated loss in Table~\ref{table:param}. We compare the SI-SDR for the $SQ$ and $SS$ scenario segments, and the power of the $QQ$ and $QS$ scenario segments. It is worth mentioning that the proposed scenario-aware differentiated loss for training is scale-sensitive, thus the extracted speech for the systems are on the same scale. When comparing between systems, the higher the better for the SI-SDR, and the lower the better for the power.

In Table~\ref{table:param}, systems 1 to 4 are trained and evaluated on the clean IEMOCAP-mix dataset, and systems 5 to 7 are trained and evaluated on the noisy IEMOCAP-mix dataset. It is seen that systems 3 and 7 achieves reasonably low power for the $QQ$ and $QS$ scenarios and high SI-SDR for the $SQ$ and $SS$ scenarios for clean and noisy dataset respectively. We therefore set $\alpha, \beta, \gamma, \delta$ to be $0.005,1,1,0.005$ respectively as in system 3 and 7 for future system evaluations. It is worth noting that the selected weights are optimized for the IEMOCAP-mix dataset, for which the total duration of the four scenario segments are shown in Table~\ref{table:iemocap}. If the total duration of the four scenario segments changes when another dataset is used, these weights need to be adjusted accordingly.

\begin{table}
    \centering
    \caption{Experiments of USEV(D) with different hyper-parameters on the clean and noisy IEMOCAP-mix validation set respectively. We report the SI-SDR (dB) for the $SQ$ and $SS$ scenario segments, and the power (dB/s) of the $QQ$ and $QS$ scenario segments. The higher the better for the SI-SDR and the lower the better for the power. N.A. represents not applicable.}
    \addtolength{\tabcolsep}{-3pt}
    \begin{tabular}{c|c |c |c |c|c|c} 
       \toprule
        \multirow{1}*{Noise}
            &\multirow{2}*{System}
            &\multirow{2}*{$\alpha$ - $\beta$ - $\gamma$ - $\delta$}
            &$QQ$    &$SQ$   &$SS$   &$QS$   \\ 
        Condition   &&&Power  &SI-SDR   &SI-SDR  &Power \\
        \midrule
        \multirow{6}*{Clean}
        &Mixture &N.A. &-79.75	&61.99   &-0.86   &8.34\\
        \cmidrule{2-7}
        &1   &0.01-1-1-0.01     &-73.80 &34.91  &6.11   &-55.88 \\
        &2   &0.01-0.1-1-0.01   &-75.91 &27.30  &6.75   &-68.71 \\
        &3   &\textbf{0.005-1-1-0.005}   &-72.84    &35.45  &6.58   &-37.87 \\
        &4   &0.001-0.1-1-0.001 &-67.87 &30.36  &7.19   &-16.02 \\
        \midrule
        \multirow{6}*{Noisy}
        &Mixture &N.A. &13.01	&3.62   &-2.12   &21.18\\
        \cmidrule{2-7}
        &5   &0.01-1-1-0.01     &-69.26 &0.15   &3.57   &-59.08  \\
        &6   &0.005-1-1-0.01    &-62.89 &1.10   &3.51   &-56.12  \\
        &7   &\textbf{0.005-1-1-0.005}   &-50.17 &4.29   &4.13   &-37.71  \\
        &8   &0.001-1-1-0.001   &-17.58 &6.74   &4.50   &-13.76  \\
       \bottomrule
    \end{tabular}
    \addtolength{\tabcolsep}{3pt}
    \label{table:param}
\end{table}

\begin{table}
    \centering
    \caption{Experiments of USEV(D) with different training strategy on the noisy IEMOCAP-mix validation set. We report the SI-SDR (dB) for the $SQ$ and $SS$ scenario segments, and the power (dB/s) of the $QQ$ and $QS$ scenario segments. The higher the better for the SI-SDR and the lower the better for the power.}
    \begin{tabular}{c|c |c |c |c|c|c} 
       \toprule
        \multirow{2}*{System}
            &\multirow{2}*{\textit{PT}}
            &\multirow{2}*{\textit{FT}}
            &$QQ$    &$SQ$   &$SS$   &$QS$   \\ 
            && &Power  &SI-SDR  &SI-SDR &Power \\
        \midrule
        Mixture &N.A. &N.A. &13.01	&3.62    &-2.12    &21.18 \\
        \midrule
        7    &\cmark   &\cmark &-50.17 &4.29     &4.13      &-37.71 \\
        \midrule
        9   &\cmark &\xmark    &-6.87  &3.28   &0.75     &6.58 \\
        10  &\xmark  &\cmark   &-65.24	&-1.75	&-0.31		&-53.67  \\
       \bottomrule
    \end{tabular}
    \label{table:training_strategy}
\end{table}

\subsubsection{Training strategy}
We evaluate the USEV(D) on the noisy IEMOCAP-mix validation set to empirically find the appropriate training strategy as shown in Table~\ref{table:training_strategy}. We compare the SI-SDR for the $SQ$ and $SS$ scenario segments, and the power of the $QQ$ and $QS$ scenario segments. The higher the better for the SI-SDR and the lower the better for the power. Systems 9 and 10 are trained and evaluated on the noisy IEMOCAP-mix dataset. System 7 in Table~\ref{table:training_strategy} is the same as system 7 in Table~\ref{table:param}. 

The USEV(D) in system 7 is pre-trained (\textit{PT}) on highly overlapped speech first (VoxCeleb2-mix dataset), and further trained (\textit{FT}) on the \textit{general speech mixtures} (IEMOCAP-mix dataset). We conduct two experiments to justify this training procedure. In system 9, USEV(D) is pre-trained on the VoxCeleb2-mix dataset, but it is not further trained on the IEMOCAP-mix dataset. It is seen that system 9 has lower SI-SDR for the $SQ$ and $SS$ scenarios, and higher power for the $QQ$ and $QS$ scenarios, as compared to system 7 which is further trained on the IEMOCAP-mix dataset. This could be due to the fact that system 9 is only trained on highly overlapped speech.

In system 10, we do not pre-train the USEV(D) on the VoxCeleb2-mix dataset, but rather train it directly on the IEMOCAP-mix dataset from scratch. System 10 significantly lags behind system 7 on the $SQ$ and $SS$ scenarios in terms of SI-SDR, where the target speaker is highly overlapped with the interference speakers or noise. We note that the $SQ$ and $SS$ scenarios are two very challenging scenarios among the four, the pre-training on highly overlapped speech is effective because it well prepares the network for the two scenarios.

\begin{table*}[ht]
    \centering
    \caption{Comparative study among variants of the SpEx+~\cite{marvin2021}, TDSE~\cite{wu2019time}, and USEV networks on IEMOCAP-mix test set. We report the average power (dB/s) of \textit{TA} speech clips, the SI-SDR (dB) for \textit{TP} speech clips by target-interference overlapping ratios, and the average SI-SDR (dB) for \textit{TP} speech clips. The higher the better for the SI-SDR and the lower the better for the power.}
    \addtolength{\tabcolsep}{-2pt}
    \begin{tabular}{c|c |c |c |c |c| c| c|c|c| c| c| c} 
       \toprule
        \multirow{1}*{Noise}
            &\multirow{2}*{Model}
            &\multirow{2}*{Reference}
            &\multirow{2}*{Extractor}
            &\multirow{2}*{Loss}
            &\textit{TA}   &0\% & (0,20]\%  &(20,40]\%  &(40,60]\% &(60,80]\%  &(80,100]\% & Average\\ 
        Condition&&&&&Power &SI-SDR  &SI-SDR  &SI-SDR  &SI-SDR  &SI-SDR  &SI-SDR &SI-SDR\\
        \midrule
        \multirow{9}*{Clean}
        &Mixture &N.A. &N.A. &N.A. &9.96  &33.41	&-0.98	&-1.32	&-1.52	&-1.53	&-1.99 &4.34\\
        \cmidrule{2-13}
        &SpEx+(S)   &\multirow{2}*{Speech}  &\multirow{2}*{TCN}  &Uniform 
            &-40.87	&12.36	&8.05	&5.36	&4.80	&3.88	&3.07	&6.25 \\
        &SpEx+(D)   &  &  &Differentiated 
            &3.13	&20.18	&10.31	&8.34	&6.19	&5.44	&4.40	&9.14 \\
        \cmidrule{2-13}
        &TDSE(S) 
            &\multirow{2}*{Visual}
            &\multirow{2}*{TCN}
            &Uniform 
            &-76.86 &23.98  &13.85  &7.58   &7.94   &6.75   &5.82   &10.99 \\
        &TDSE(D)    & &   &Differentiated
            &-22.82 &29.80  &15.72  &9.70   &8.29   &6.95   &6.13   &12.76 \\
        \cmidrule{2-13}
        &USEV(S)        &\multirow{2}*{Visual} &\multirow{2}*{DPRNN} &Uniform  
            &-77.48 &30.53  &15.23  &7.14   &7.37   &5.96   &5.46   &11.95 \\
        &USEV(D)       &&&Differentiated
            &-43.00 &33.83  &15.20  &9.91   &8.06   &6.89   &5.95   &13.31 \\
        \midrule
        \multirow{9}*{Noisy}
        &Mixture &N.A. &N.A. &N.A. &22.93    &1.9    &-2.53  &-2.47  &-2.76  &-2.79  &-2.91  &-1.93\\
        \cmidrule{2-13}
        &SpEx+(S)   &\multirow{2}*{Speech}  &\multirow{2}*{TCN}  &Uniform 
            &-32.27	&1.76	&0.11	&1.84	&1.67	&0.85	&1.13	&1.22 \\
        &SpEx+(D)   &  &  &Differentiated 
            &2.93	&6.66	&0.96	&3.02	&2.43	&2.16	&1.54	&2.79 \\
        \cmidrule{2-13}
        &TDSE(S) 
            &\multirow{2}*{Visual}
            &\multirow{2}*{TCN}
            &Uniform 
            &-75.46 &4.92   &3.79   &3.79   &4.56   &4.17   &3.58   &4.14 \\
        &TDSE(D)    & &   &Differentiated
            &-32.93 &8.56   &6.64   &6.61   &5.80   &4.90   &4.29   &6.13 \\
        \cmidrule{2-13}
        &USEV(S)        &\multirow{2}*{Visual} &\multirow{2}*{DPRNN}&Uniform  
            &-74.04 &5.01   &3.16   &3.78   &4.11   &3.69   &3.36   &3.85 \\
        &USEV(D)       &&&Differentiated
            &-22.23 &8.97   &6.05   &6.22   &5.30   &4.46   &4.14   &5.86 \\
       \bottomrule
    \end{tabular}
    \addtolength{\tabcolsep}{2pt}
    \label{table:compare_overlap}
\end{table*}

\begin{table*}[ht]
    \centering
    \caption{Comparative study among variants of the SpEx+~\cite{marvin2021}, the TDSE~\cite{wu2019time}, and USEV networks on IEMOCAP-mix test set. We report the SI-SDR for the $SQ$ and $SS$ scenario segments, and the power of the $QQ$ and $QS$ scenario segments. The higher the better for the SI-SDR and the lower the better for the power.}
    \begin{tabular}{c|c|c |c |c |c |c| c| c|c} 
      \toprule
        \multirow{1}*{Noise}
            &\multirow{2}*{Model}
            &\multirow{2}*{Reference}
            &\multirow{2}*{Extractor}
            &\multirow{2}*{Loss}
            &$QQ$    &$SQ$   &$SS$   &$QS$ &\multirow{1}*{Parameters}\\ 
        Condition&&&& &Power     &SI-SDR  &SI-SDR  &Power    &(million)\\
        \midrule
        \multirow{9}*{Clean}
        &Mixture &N.A. &N.A. &N.A.   &-80    &63.02  &-0.8573    &10.23   &-\\
        \cmidrule{2-10}
        &SpEx+(S)  &\multirow{2}*{Speech}  &\multirow{2}*{TCN}  &Uniform 
                            &-45.56 &14.56  &3.50   &-17.15 &\multirow{2}*{11.9}\\
        &SpEx+(D)    &&&Differentiated
                            &-42.25 &22.95  &5.09   &-14.65 & \\
        \cmidrule{2-10}
        &TDSE(S) &\multirow{2}*{Visual}  &\multirow{2}*{TCN} &Uniform 
                            &-65.17 &23.42  &5.75   &-29.37 &\multirow{2}*{22.1}\\
        &TDSE(D)     &&&Differentiated
                            &-69.29 &33.88 &6.33    &-30.24&\\
        \cmidrule{2-10}
        &USEV(S)     &\multirow{2}*{Visual} &\multirow{2}*{DPRNN}&Uniform  
                            &-65.29 &24.26  &5.35   &-29.17 &\multirow{2}*{15.3} \\
        &USEV(D)     &&&Differentiated
                            &-71.87 &35.40  &6.19   &-35.58 & \\
        \midrule
        \multirow{9}*{Noisy}
        &Mixture &N.A. &N.A. &N.A.   &13.33   &3.88   &-2.05  &21.30  &-\\
        \cmidrule{2-10}
        &SpEx+(S)  &\multirow{2}*{Speech}  &\multirow{2}*{TCN}  &Uniform 
                            &-27.45 &2.69   &1.08   &-13.66 &\multirow{2}*{11.9}\\
        &SpEx+(D)    &&&Differentiated
                            &-27.10 &4.87   &2.07   &-7.91& \\
        \cmidrule{2-10}
        &TDSE(S) &\multirow{2}*{Visual}  &\multirow{2}*{TCN} &Uniform 
                            &-35.91 &4.68   &3.03   &-26.12 &\multirow{2}*{22.1}\\
        &TDSE(D)     &&&Differentiated
                            &-59.38 &3.27   &4.04   &-40.14&\\
        \cmidrule{2-10}
        &USEV(S)     &\multirow{2}*{Visual} &\multirow{2}*{DPRNN}&Uniform  
                            &-34.37 &4.45   &2.76   &-25.28 &\multirow{2}*{15.3} \\
        &USEV(D)     &&&Differentiated
                            &-50.04 &4.85   &3.83   &-34.00 & \\
      \bottomrule
    \end{tabular}
    \label{table:compare_scenario}
\end{table*}

\begin{figure*}
\begin{minipage}[t]{.48\linewidth}
  \centering
  \centerline{\includegraphics[width=8.0cm, height=4.0cm]{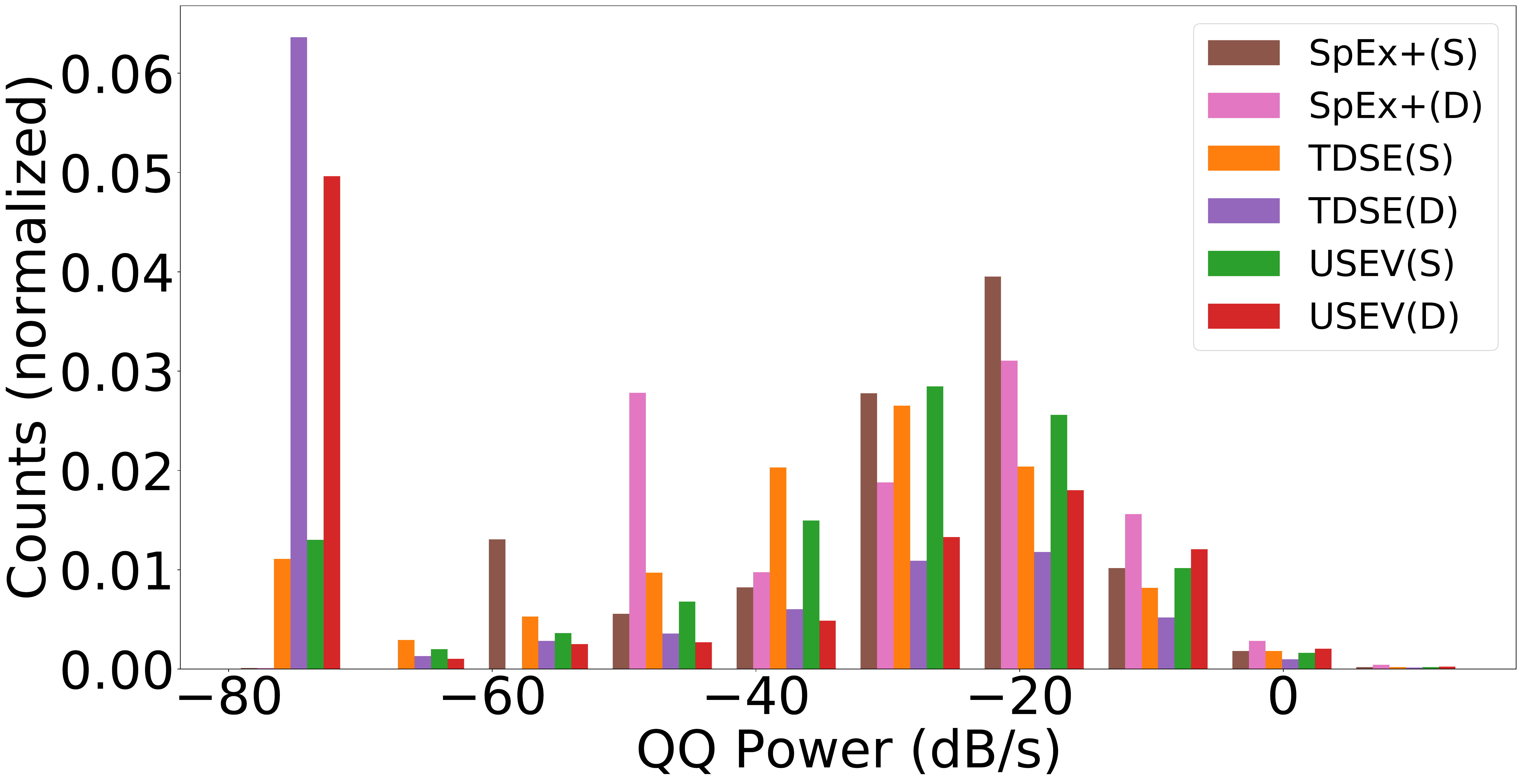}}
\end{minipage}
\hfill
\begin{minipage}[t]{0.48\linewidth}
  \centering
  \centerline{\includegraphics[width=8.0cm, height=4.0cm]{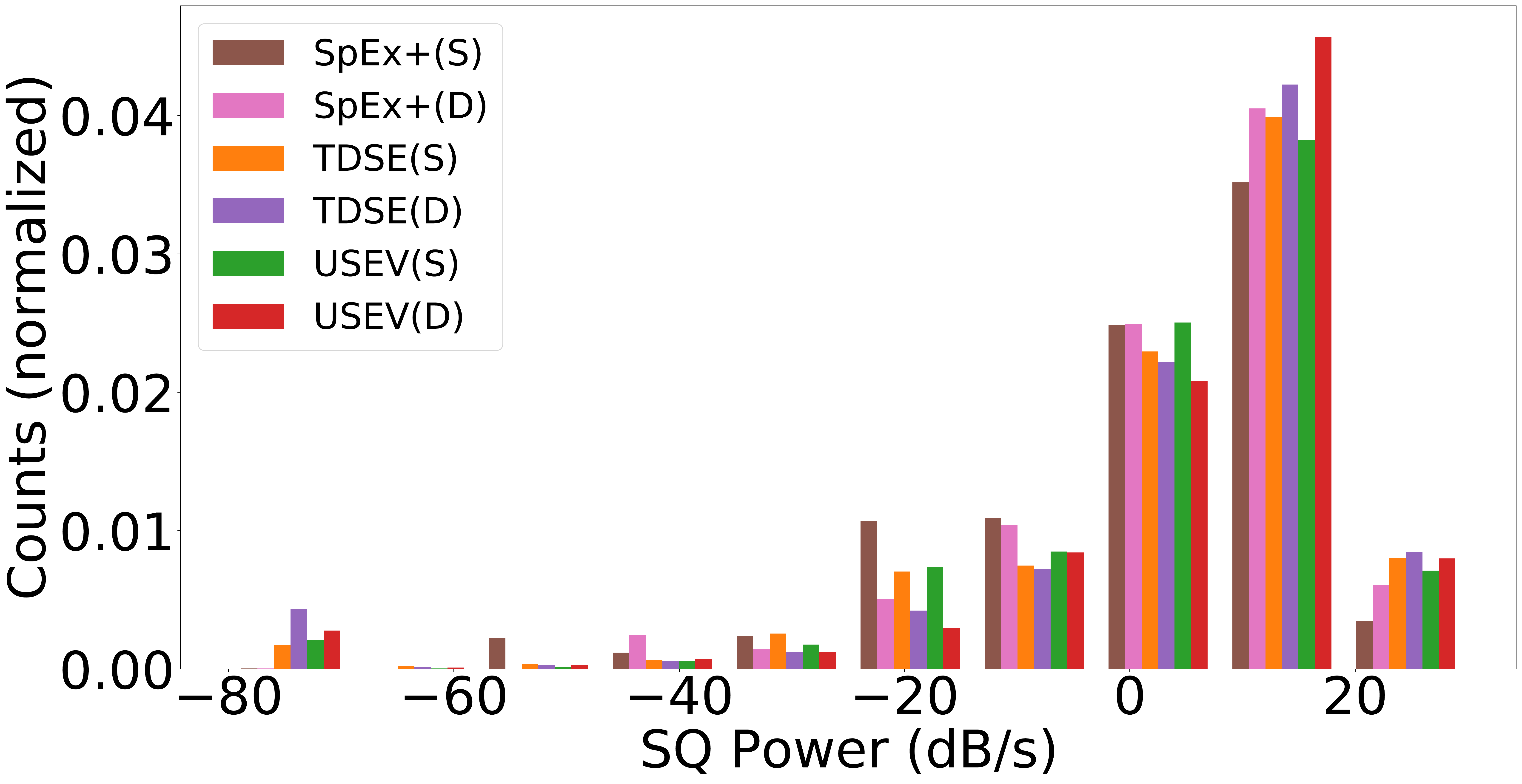}}
\end{minipage}
\vskip\baselineskip
\begin{minipage}[t]{0.48\linewidth}
  \centering
  \centerline{\includegraphics[width=8.0cm, height=4.0cm]{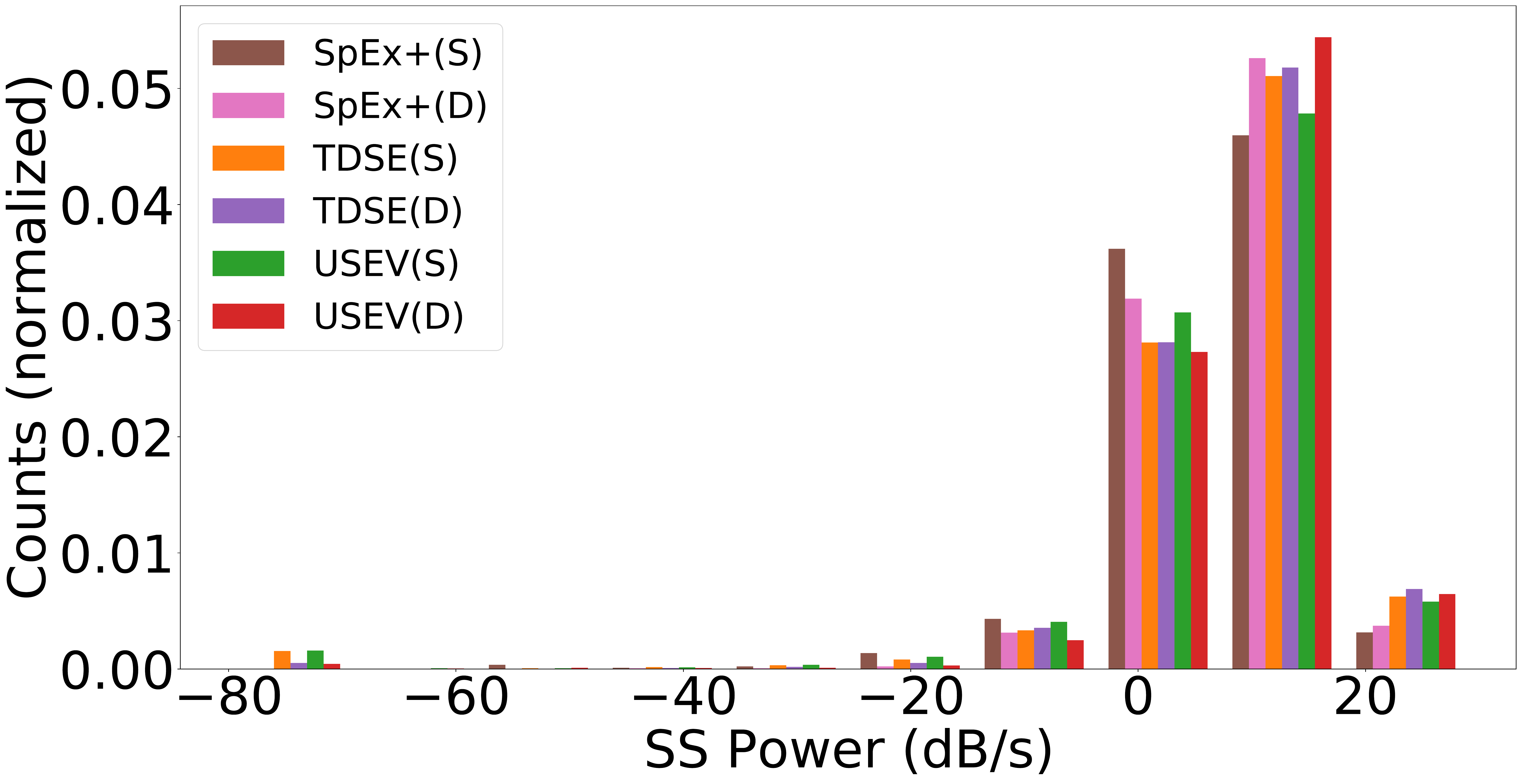}}
\end{minipage}
\hfill
\begin{minipage}[t]{0.48\linewidth}
  \centering
  \centerline{\includegraphics[width=8.0cm, height=4.0cm]{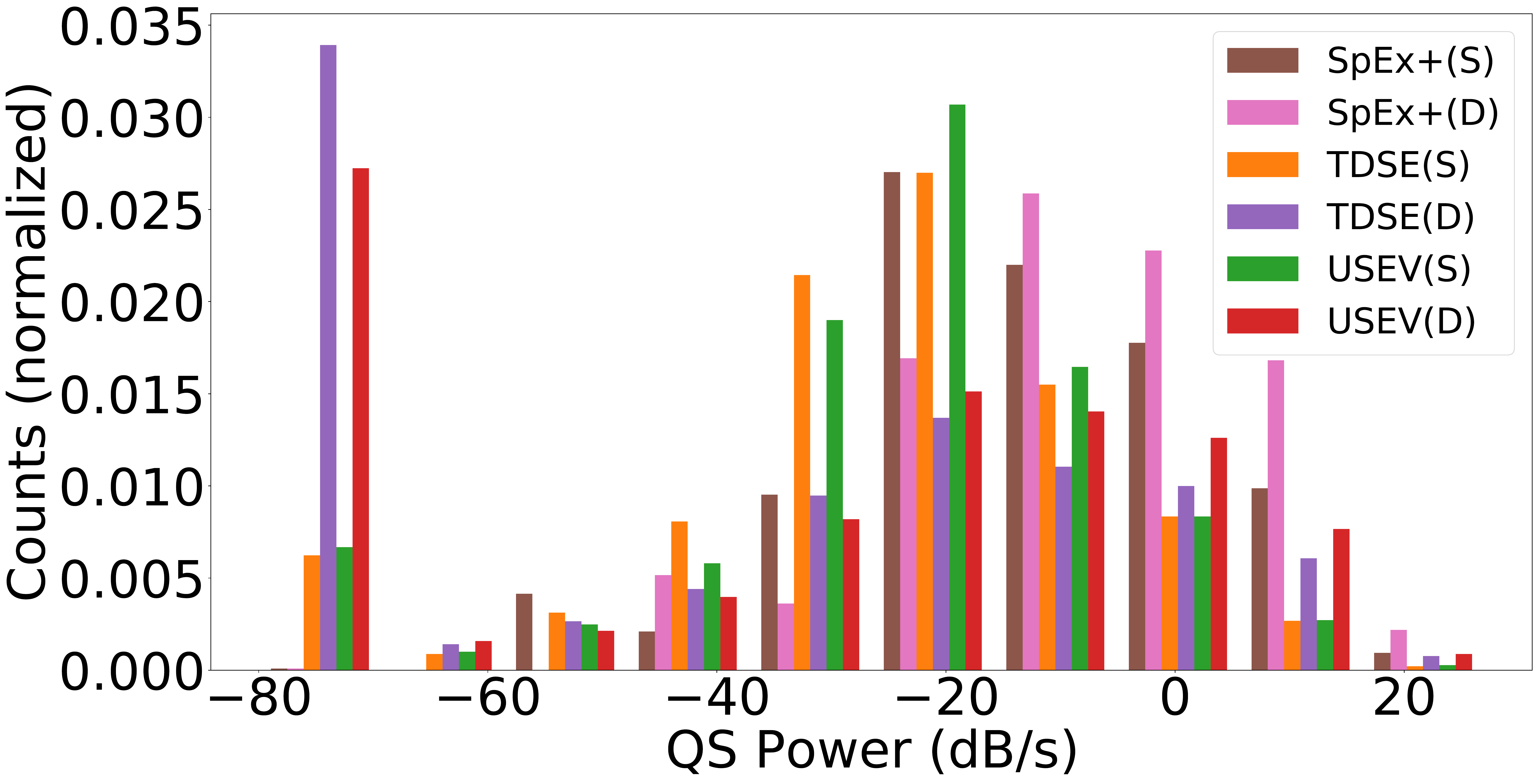}}
\end{minipage}
\caption{The histogram of the power by the SpEx+(S), SpEx+(D), TDSE(S), TDSE(D), USEV(S), and USEV(D) on the noisy IEMOCAP-mix test sets. The lower the better for the power of the $QQ$ and $QS$ scenario, and the power of the $SQ$ and $SS$ scenario are expected to be similar to that of the clean target speech, which is on average 10 dB/s.}\medskip
\label{fig:power_plot}
\end{figure*}

\subsection{Comparative studies}
\label{sec:results_compare}

The SpEx+(S), TDSE(S), and USEV(S) adopt the uniform loss as the training objective, while the SpEx+(D), TDSE(D), and USEV(D) adopt the differentiated loss. For a fair comparison between the two loss functions, we report the performance on an entire speech clip with a single evaluation metric, i.e., SI-SDR for \textit{TP} speech lips or power for \textit{TA} speech clips in Table~\ref{table:compare_overlap}. 

In Table~\ref{table:compare_overlap}, we first categorize the speech clips into two categories: i) \textit{TA} speech clips: speech clips with the absent target speaker. ii) \textit{TP} speech clips: speech clips with the present target speaker. We report the average power of \textit{TA} speech clips, the lower the better. For \textit{TP} speech clips, we report the SI-SDR by their target-interference speaker overlapping ratios. The average SI-SDR is also reported in the last column, the higher the better for the SI-SDR.

In addition, we report the results using differentiated metrics in terms of $QQ, SQ, SS, QS$ scenarios in Table~\ref{table:compare_scenario}, to examine the performance over individual scenarios. The models reported in Table~\ref{table:compare_overlap} and Table~\ref{table:compare_scenario} are both trained on the IEMOCAP-mix dataset, but are evaluated using different metrics, i.e., a single SI-SDR or power over an entire speech clip in Table~\ref{table:compare_overlap}, and differentiated metrics for every scenario segment in Table~\ref{table:compare_scenario}.

\subsubsection{Visual vs audio cue}
We compare the use of visemes with the use of a pre-recorded speech as the auxiliary reference on the IEMOCAP-mix test set. The SpEx+ network employs a pre-recorded speech as the audio cue, while the TDSE and USEV networks employ the visemes as the visual cue. The pre-recorded speech used for the SpEx+ is the target's other clean speech utterance that is not present in the speech mixture, the reference utterance is about $5$ seconds long on average.

The SpEx+(S), TDSE(S), and USEV(S) are trained using the uniform loss. As shown in Table~\ref{table:compare_overlap}, for both clean and noisy conditions, the TDSE(S) and USEV(S) outperform the SpEx+(S) in terms of average SI-SDR, and the \textit{TA} power. As shown in Table~\ref{table:compare_scenario}, for both clean and noisy conditions, the TDSE(S) and USEV(S) outperform the SpEx+(S) for all scenarios. Results show that when the uniform loss is used, the visemes outperform the pre-recorded speech as the auxiliary reference.

The SpEx+(D), TDSE(D), and USEV(D) are trained using the differentiated loss. The weights of the differentiated loss are selected the same as system 7 in Table~\ref{table:param}. As shown in Table~\ref{table:compare_overlap}, for both clean and noisy conditions, the TDSE(D) and USEV(D) consistently outperform the SpEx+(D) in terms of average SI-SDR, and the \textit{TA} power. As shown in Table~\ref{table:compare_scenario}, for both clean and noisy conditions, the USEV(D) and TDSE(D) outperform the SpEx+(D) for all scenarios, except for the $SQ$ scenario under noisy conditions. Results show that when the differentiated loss is used, the visemes outperform the pre-recorded speech as the auxiliary reference.

\subsubsection{Differentiated loss vs uniform loss}
We compare the use of differentiated loss with the uniform loss on the baselines and our proposed USEV network in disentangling the \textit{general speech mixture} clips from the IEMOCAP-mix test set.

As shown in Table~\ref{table:compare_overlap}, for both clean and noisy conditions, the SpEx+(D) performs better in terms of SI-SDR evaluation, but the SpEx+(S) performs better in terms of \textit{TA} power. Further look into Table~\ref{table:compare_scenario}, for both clean and noisy conditions, the SpEx+(D) performs better for the $SQ$ and $SS$ scenarios, but the SpEx+(S) has lower power for the $QQ$ and $QS$ scenarios. For the SpEx+ network, the uniform loss focuses more on muting the network for inactive target speaker scenarios, while the differentiated loss focuses more on the signal extraction quality for active target speaker scenarios, this could be caused by the large weights used for the $SQ$ and $SS$ scenarios used by the differentiated loss function.

As shown in Table~\ref{table:compare_overlap}, for both clean and noisy conditions, the TDSE(D) outperforms the TDSE(S) in terms of the SI-SDR evaluation, but the TDSE(S) performs better for \textit{TA} power. As shown in Table~\ref{table:compare_scenario}, for both clean and noisy conditions, the TDSE(D) outperforms the TDSE(S) for scenarios, except for the $SQ$ scenario under noisy conditions.

As shown in Table~\ref{table:compare_overlap}, for both clean and noisy conditions, the USEV(D) outperforms USEV(S) in terms of the SI-SDR evaluation, but the USEV(S) performs better for \textit{TA} power. As shown in Table~\ref{table:compare_scenario}, for both clean and noisy conditions, the USEV(D) outperforms USEV(S) for all scenarios. 

In Table~\ref{table:compare_overlap}, the uniform loss consistently performs better than the differentiated loss in terms of power for \textit{TA} speech clips. This is because the uniform loss puts a high weight of 1 on the \textit{TA} speech clips during training, as opposed to 0.005 for the differentiated loss ($QQ$ and $QS$). It is worth noting that the \textit{TA} speech clips are only a subset of all $QQ$ and $QS$ scenario segments. If we compare the overall power of the $QQ$ and $QS$ scenarios in Table~\ref{table:compare_scenario}, the differentiated loss consistently outperforms the uniform loss in terms of power for TDSE and USEV models.

We also present the power distribution of the models on the noisy IEMOCAP-mix test sets in Fig.~\ref{fig:power_plot}, It is seen that for the $SQ$ and $SS$ scenarios in which the target speaker is active, all models perform similar, with most of the extracted speech signals have a power value that is close to that of the clean target speech, which is on average 10 dB/s. For the $QQ$ and $QS$ scenarios in which the target speaker is inactive, it is seen that the USEV(D) and TDSE(D) have more samples with lower power compared to the USEV(S) and TDSE(S), showing that the differentiated loss outperforms the uniform loss for audio-visual models.

\begin{table*}[ht]
    \centering
    \caption{Comparative studies among variants of the TDSE~\cite{wu2019time}, and USEV networks on IEMOCAP-mix test set for a visual cue with occlusion. The values reported are the average over all visual occlusion settings in its group, for models that are trained on IEMOCAP-mix training data with and without occlusion (Occl.).}
    \begin{tabular}{c|c |c |c |c |c| c| c|c|c| c| c} 
       \toprule
        \multirow{2}*{Model}
            &\multirow{2}*{Extractor}
            &\multirow{2}*{Loss}
            &Train
            &\textit{TA}   &0\% & (0,20]\%  &(20,40]\%  &(40,60]\% &(60,80]\%  &(80,100]\% & Average\\ 
        &&&Occl.    &Power  &SI-SDR  &SI-SDR  &SI-SDR  &SI-SDR  &SI-SDR  &SI-SDR&SI-SDR \\
        
        \midrule
        Mixture &N.A. &N.A. &N.A. &9.96 &33.41	&-0.98	&-1.32	&-1.52	&-1.53	&-1.99 &4.34\\
        
        \midrule
        TDSE(S) &TCN &Uniform &\multirow{4}*{\xmark}
            &-76.98	&-11.50	&-3.28	&-14.85	&-20.36	&-16.09	&-12.53	&-13.10\\
        TDSE(D) &TCN &Differentiated  &
            &-12.72	&15.19	&6.82	&3.22	&0.54	&-0.16	&-0.51	&4.18\\
        USEV(S)   &DPRNN&Uniform  &
            &-75.34	&-8.88	&-6.97	&-20.40	&-25.67	&-21.76	&-16.96	&-16.77\\
        USEV(D)        &DPRNN&Differentiated  &
            &-29.03	&15.88	&5.50	&2.62	&0.04	&-0.44	&-0.58	&3.84\\
        \midrule
        TDSE(S) &TCN &Uniform &\multirow{4}*{\cmark} 
            &-61.57	&23.48	&13.14	&7.40	&5.90	&5.63	&4.92	&10.08\\
        
        TDSE(D) &TCN &Differentiated  &
            &5.24	&25.56	&12.87	&8.00	&6.10	&5.55	&4.60	&10.45\\
        USEV(S)   &DPRNN    &Uniform  &
            &-67.28	&24.40	&12.08	&4.07	&4.48	&3.63	&3.77	&8.74\\
        USEV(D)   &DPRNN   &Differentiated  &
            &-6.03	&31.11	&12.45	&7.57	&5.73	&5.00	&4.37	&11.04\\
       \bottomrule
    \end{tabular}
    \label{table:occlusion}
\end{table*}

\begin{figure*}
\begin{minipage}[t]{.24\linewidth}
  \centering
  \centerline{\includegraphics[width=4.5cm, height=3.0cm]{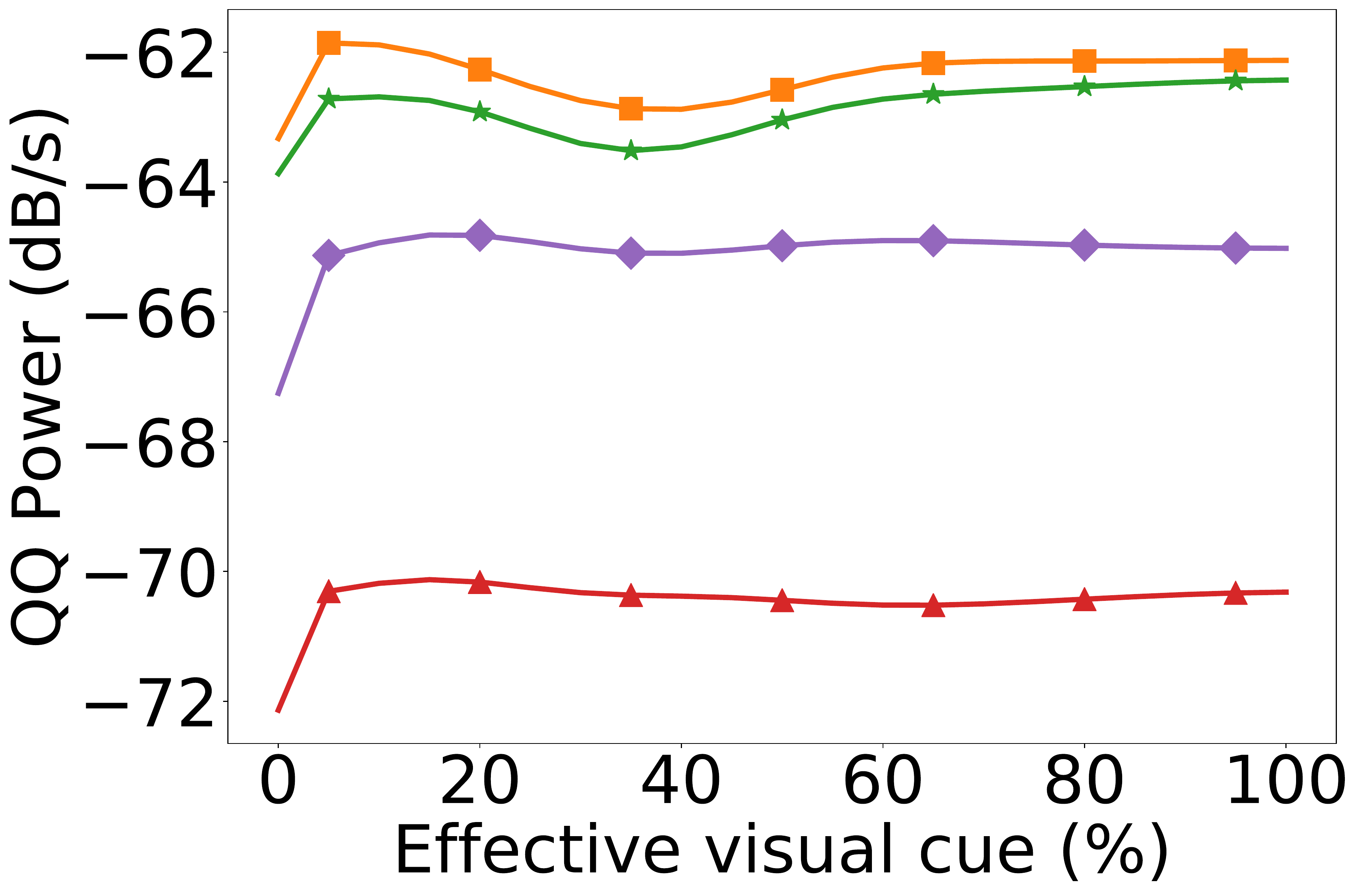}}
  \centerline{\scalebox{0.8}{(a)}}\medskip
\end{minipage}
\hfill
\begin{minipage}[t]{0.24\linewidth}
  \centering
  \centerline{\includegraphics[width=4.5cm, height=3.0cm]{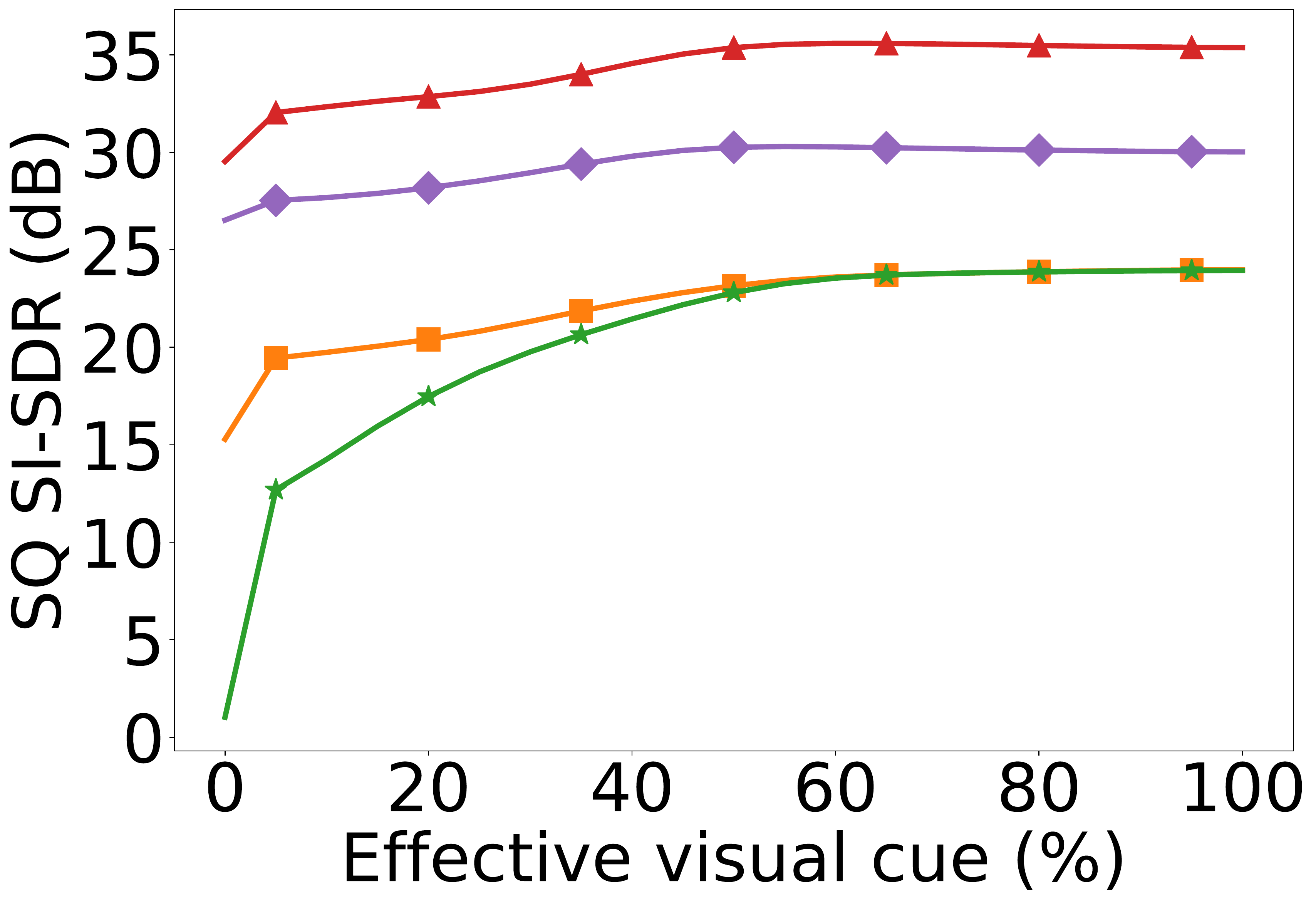}}
  \centerline{\scalebox{0.8}{(b)}}\medskip
\end{minipage}
\hfill
\begin{minipage}[t]{0.24\linewidth}
  \centering
  \centerline{\includegraphics[width=4.5cm, height=3.0cm]{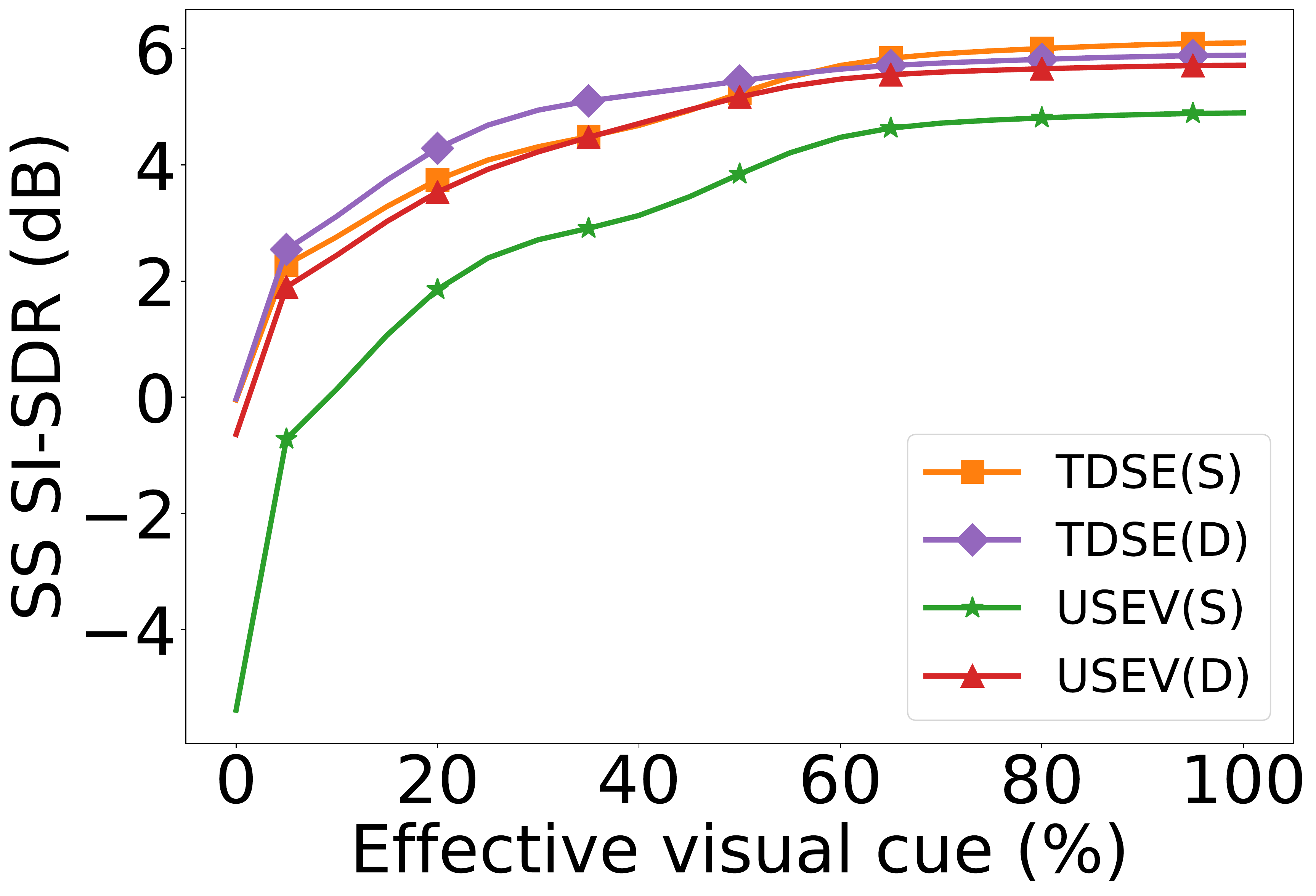}}
  \centerline{\scalebox{0.8}{(c)}}\medskip
\end{minipage}
\hfill
\begin{minipage}[t]{0.24\linewidth}
  \centering
  \centerline{\includegraphics[width=4.5cm, height=3.0cm]{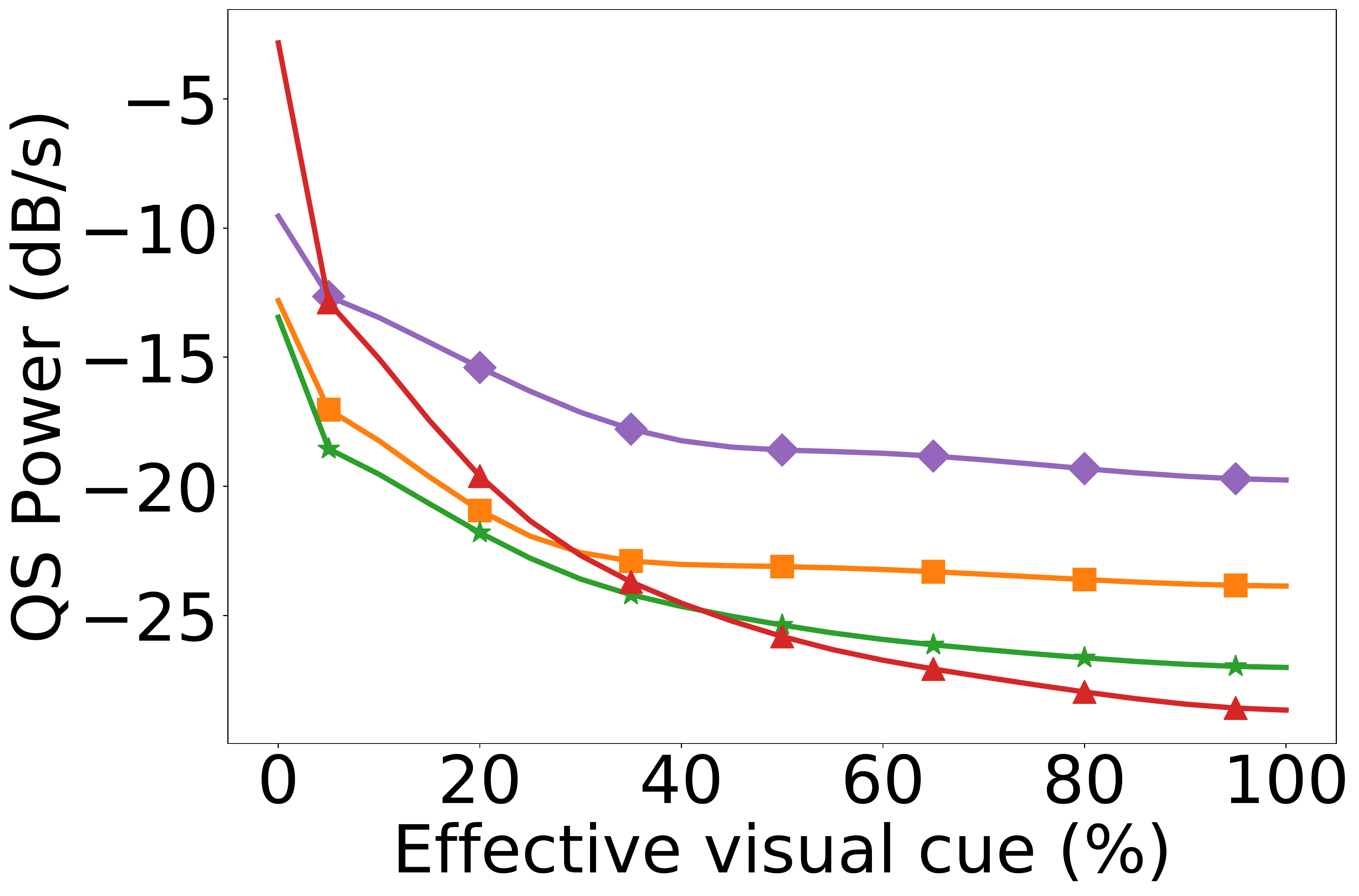}}
  \centerline{\scalebox{0.8}{(d)}}\medskip
\end{minipage}
\caption{Comparative studies among the TDSE(S), TDSE(D), USEV(S), and USEV(D) with partially occluded visual reference. The models are trained on the IEMOCAP-mix training data with visual occlusion.  We report the performance in $QQ, SQ, SS, QS$ scenarios separately as a function of the effective visual cue, i.e., the ratio of the effective visual duration to the total length of the speech clip. For each model, the lower the better for power in (a) $QQ$ scenario and (d) $QS$ scenario, and the higher the better for SI-SDR in (b) $SQ$ scenario and (c) $SS$ scenario.}\medskip
\label{fig:occl_plots}
\end{figure*}

\subsubsection{DPRNN vs TCN}
We compare the DPRNN with TCN in the speaker extractor when the visual cue is employed as the auxiliary reference. The TDSE network employs a repeated stack of TCN in the speaker extractor to estimate the receptive mask, while the USEV network employs a repeated DPRNN block to estimate the receptive mask.

As shown in Table~\ref{table:compare_overlap}, when the uniform loss is used for network training, the USEV(S) outperforms the TDSE(S) for clean conditions, but the TDSE(S) outperforms the USEV(S) for noisy conditions. As shown in Table~\ref{table:compare_scenario}, for clean conditions, the USEV(S) performs better for the $QQ$ and $SQ$ scenarios, but the TDSE(S) performs better for the $SS$ and $QS$ scenarios, for noisy condition, the TDSE(S) performs better than the USEV(S) for all scenarios. 

As shown in Table~\ref{table:compare_overlap}, when the differentiated loss is used for network training, the USEV(S) outperforms the TDSE(D) for clean conditions, but the TDSE(D) outperforms the USEV(D) for noisy conditions. As shown in Table~\ref{table:compare_scenario}, the USEV(D) outperforms the TDSE(D) in all scenarios except for the $SS$ scenario under clean condition, the TDSE(D) outperforms the USEV(D) in all scenarios except for the $SQ$ scenario under noisy condition.

The USEV tends to perform better when the overlapping ratio is small and the TDSE tends to perform better when the overlapping ratio is large. Overall, the USEV and TDSE networks achieve comparable results on both clean and noisy IEMOCAP-mix test sets. However, the USEV network has $6.8$ million fewer parameters compared with the TDSE network due to the smaller size of the DPRNN blocks as shown in Table~\ref{table:compare_scenario}. In addition, when the kernel size $L$ in the speech encoder of the two networks is set the same, which is $40$ in this paper, the training of the USEV network is $4$ times faster than that of the TDSE network. The TDSE and USEV network have more network parameters compared with the SpEx+ network due to the large visual encoder module. The no. of network parameters reported for the TDSE and USEV networks do not include the face detection and tracking module.

\subsection{Effect of visual occlusion}
\label{sec:results_visual}

When face detection and tracking algorithms fail to detect the target speaker or the lip is occluded for some reason, the visual cue is absent or incomplete. We simulate the above scenario, referred to as visual occlusion, by setting a random duration of the lip images sequence to zero signals for each speech clip in the clean IEMOCAP-mix dataset, while keeping the audio signal intact. We study the impact of such visual occlusion on the audio-visual models, namely the TDSE(S), TDSE(D), USEV(S), and USEV(D), and present their results in Table~\ref{table:occlusion}.

The models are first pre-trained on the VoxCeleb2-mix dataset, where there is no occlusion in the training data. In the training stage with the IEMOCAP-mix dataset, if the occlusion data are not involved (Train Occl. \xmark), none of the models perform well on occlusion evaluations. If the occlusion data are involved in the training stage on the IEMOCAP-mix dataset (Train Occl. \cmark), all models improve, with USEV(D) achieving the highest average SI-SDR value of 11.04 dB. The USEV tends to perform better when the overlapping ratio is small and the TDSE tends to perform better when the overlapping ratio is large, overall they achieve comparable performance for the visual occlusion analysis.

We also evaluate the models with the different effective visual cue and report the results in Fig.~\ref{fig:occl_plots} for the four scenarios, to observe the trend of model performance with the different effective visual cue. We define the effective visual cue (\%) as the ratio of the non-occluded visual duration to the total length of the speech clip. The average performance for each 5\% interval of the effective visual cue is plotted, e.g, the first point is the average performance for 0-5\% effective visual cue, and the last point is 96-100\% effective visual cue. It is observed that the performance of $SQ, SS, QS$ scenarios improves as the effective visual duration increases. When the effective visual cue reaches 20\%, the USEV(D) performs reasonably well across all scenarios, when it goes beyond 50\%, the USEV(D)'s performance starts to saturate.

\section{Conclusion}
\label{sec:conclusion}
In this paper, we propose a categorization of a \textit{general speech mixture} into four target-interference speaker pairing scenarios namely $QQ, SQ, SS, QS$. We also proposed a universal speaker extraction network with a visual cue to disentangle the \textit{general speech mixture} with a scenario-aware differentiated loss function. The experiments show that the proposed differentiated loss function is effective. This paper marks an important step towards solving a realistic \textit{cocktail party} problem.


%





\ifCLASSOPTIONcaptionsoff
  \newpage
\fi





\bibliographystyle{IEEEtran}
\bibliography{IEEEabrv,Bibliography}

\vfill


\end{document}